\documentclass[sn-apa]{sn-jnl}

\usepackage{graphics,graphicx}
\usepackage{wrapfig}
\usepackage{natbib}
\usepackage{amsmath}
\usepackage{amssymb}
\usepackage{tabularx}
\usepackage{enumitem}
\usepackage{multicol}
\usepackage[T1]{fontenc}

\jyear{2022}%

\theoremstyle{thmstyleone}%
\theoremstyle{thmstyletwo}%

\theoremstyle{thmstylethree}%

\begin{document}

\title[Transition disks]{Transition disks: the observational revolution from SEDs to imaging}


\author*[1]{\fnm{Nienke} \sur{van der Marel}}\email{nmarel@strw.leidenuniv.nl}

\affil*[1]{\orgdiv{Leiden Observatory}, \orgname{Leiden University}, \orgaddress{\street{P.O. Box 9513}, \city{Leiden}, \postcode{2333 CA}, \country{the Netherlands}}}


\abstract{
Protoplanetary disks surrounding young stars are the birth place of planets. Of particular interest are the transition disks with large inner dust cavities of tens of au, hinting at the presence of massive companions. These cavities were first recognized by a deficit in their Spectral Energy Distribution (SED), later confirmed by millimeter interferometry observations. The Atacama Large Millimeter/submillimeter Array (ALMA) has truly revolutionized the field of spatially resolved imaging of protoplanetary disks in both dust and gas, providing important hints for the origin of gaps and cavities. At the same time, new types of substructures have been revealed. Also infrared observations show a large range of substructures both in resolved imaging, interferometry and spectroscopy. Since the last review paper of transition disks in Protostars and Planets VI, a huge amount of data has been taken, which led to numerous new insights in the origin of transition disks. In this review I will summarize the observational efforts from the past decade, compare their insights with the predictions from SED modeling, analyze the properties of the transition disk population and discuss their role in general disk evolution.}


\maketitle

\section{Introduction}
\label{sec:intro}
The study of protoplanetary disks and their dissipation processes is one of the most active fields in today's astronomy \citep[e.g.][]{WilliamsCieza2011,Testi2014,Ercolano2017,Andrews2020,Manara2022}. The investigation of transition disks, loosely defined as disks with an inner clearing in their dust distribution \citep[e.g.][]{Espaillat2014}, is a fundamental component of these studies, as these disks were historically the only targets with direct evidence for clearing with respect to the younger, full circumstellar disks. Early evidence for clearing of the dust was the deficit in the mid infrared excess of the Spectral Energy Distribution (SED), whereas the excess would be high again at longer wavelengths, comparable to that of full protoplanetary disks \citep[e.g.][]{Calvet2002}. The clearing is also referred to as a `hole' or `cavity'. What makes these disks particularly intriguing is that they still maintain high accretion rates, comparable to those of `full' disks, indicating that some material is still flowing through the gap towards the star. As transition disks were identified long before the rise of spatially resolved imaging with e.g. millimeter interferometry, scattered light imaging and (more recently) optical interferometry, it is important to keep in mind that early studies and transition disk identifications may  actually represent a large range of actual disk morphologies and physical processes responsible for the clearing. The literature has become cluttered with a large range of definitions and classifications within the term `transition disk', both in the observational and the disk modeling perspectives. The most recent reviews that were specifically focused on transition disks date from Protostars \& Planets VI in 2014 \citep{Espaillat2014} as well as a PASA review in 2016 \citep{Casassus2016} which pre-dates the majority of the high-resolution millimeter studies with the Atacama Large Millimeter/submillimeter Array (ALMA) \citep[e.g.][]{Pinilla2018tds,Francis2020} and the efforts on both high contrast imaging of embedded protoplanets and scattered light imaging with optical/infrared (OIR) facilities such as VLT/SPHERE, Gemini/GPI and Subaru/SCExAO \citep{Benisty2022}. Therefore, this new review attempts to summarize the different contributions to the field over the last 10 years, discuss the implications of the different definitions of transition disks and compare the transition disks as identified from SEDs and (millimeter) imaging. The focus of the latter will be primarily on millimeter dust continuum images, as the morphologies and relevant disk processes of scattered light imaging was recently reviewed in \citep{Benisty2022}. 

\begin{figure}[!ht]
\centering
\includegraphics[width=0.9\textwidth]{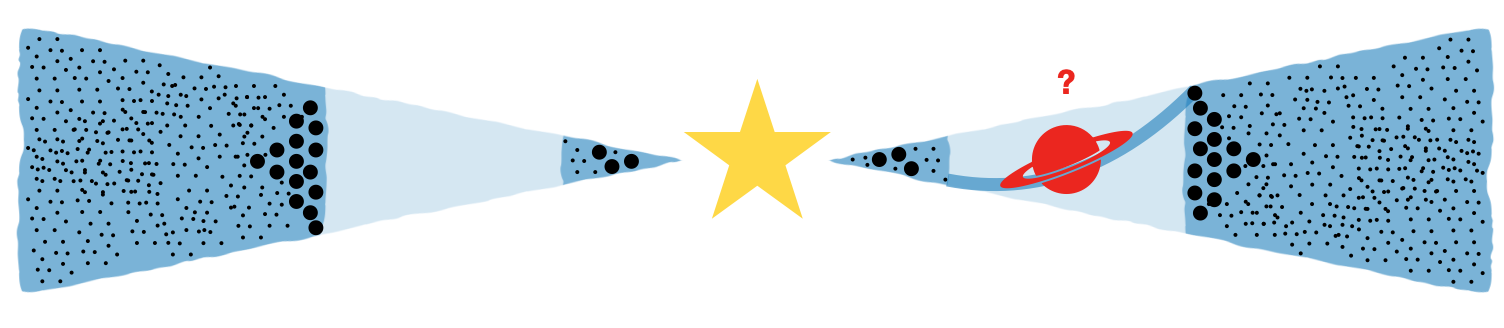}
\caption{Sketch of a transition disk structure of a large cavity transition disk where millimeter dust is trapped at the cavity edge. The gap may be caused by a giant planet but material is still flowing through the gap as evidenced from the high accretion rates.}
\label{fig:sketch}
\end{figure}

\begin{figure}[!ht]
\centering
\includegraphics[width=0.9\textwidth]{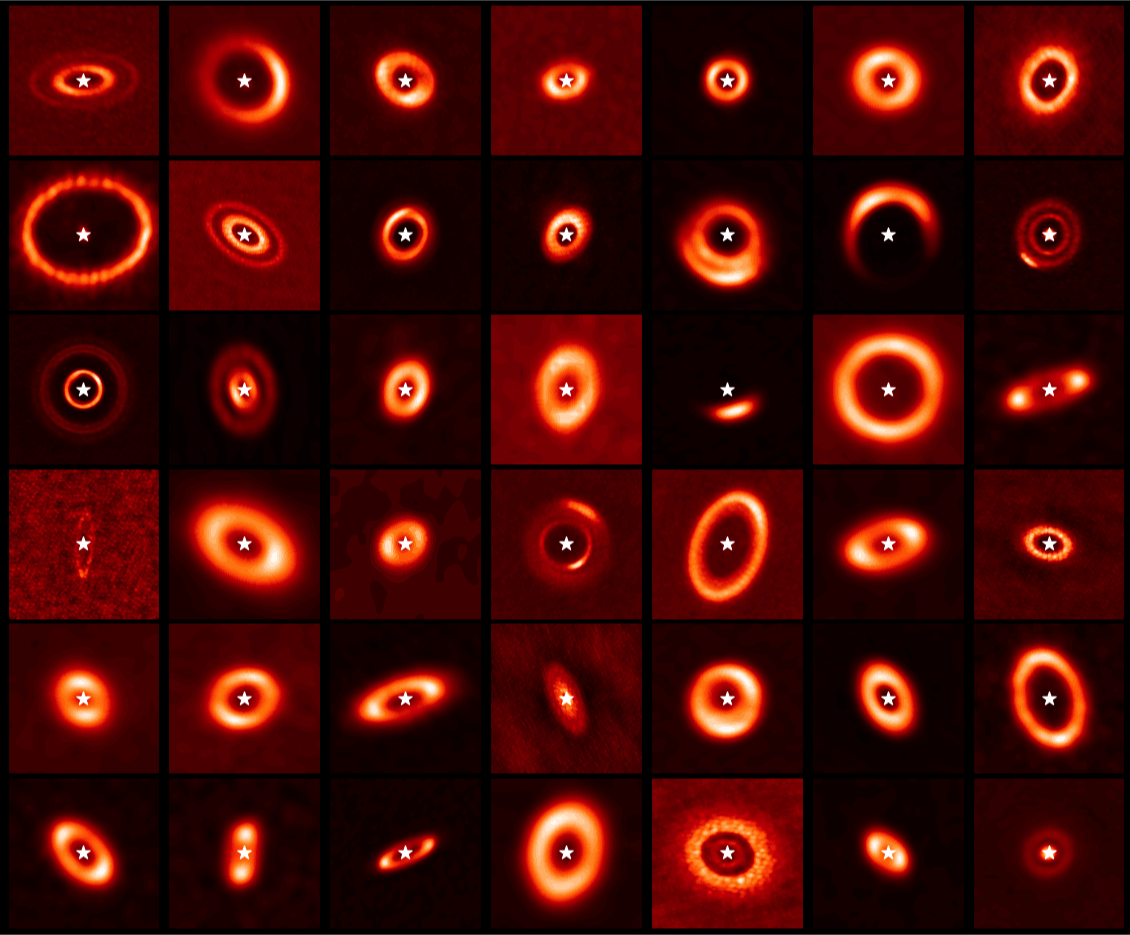}
\caption{Gallery of continuum emission of transition disks imaged with ALMA, showing the large diversity of substructures.} 
\label{fig:gallery}
\end{figure}

Historically, the term `transition disk' (sometimes also referred to as `transitional disk') originates from the first discovery of 8 young stars with infrared excess at $\lambda>10\mu$m and the absence of excess at $\lambda<10\mu$m in a large sample of 83 pre-main sequence stars in Taurus with near-infrared (\emph{J, H, K, L}-band) and \emph{IRAS} (25, 60, 100 and 160 $\mu$m) photometry \citep{Strom1989}. The authors define these disks as being "in transition", possibly due to "disk clearing" where the "inner disk regions $r<0.1$ au are relatively devoid of distributed matter". The occurrence of these disks of approximately 10\% of the sample is regarded as a measure of the "transition time" from massive, optically thick to lower-mass, more tenuous structures \citep[e.g.][]{KenyonHartmann1995,Simon1995}, a fraction that is recovered in more recent studies as well \citep[e.g.][]{Ercolano2017}. Interestingly, the 8 identified transition disks by Strom et al. are all but one marked as weak-line T Tauri star, i.e. a non-accreting young star, which would be consistent with these disks being caught in the act of dissipation, making them indeed `transitional'. However, a closer inspection of these disks with \emph{Spitzer} and \emph{Herschel} photometry shows that the \emph{IRAS} photometric points of several of these disks are confused with other emission due to the large beam size and thus not actual transition disks as discovered in later decades \citep[e.g.][]{Ribas2016}.

New infrared facilities such as \emph{ISO} and \emph{Spitzer} led to the discovery of many more transition disks with an infrared deficit, e.g. a 4 au cavity in TW~Hya  \citep{Calvet2002}, and larger cavities $>$10 au in e.g. HD100546 \citep{Bouwman2003}, DM~Tau and GM~Aur \citep{Calvet2005}, CoKu Tau/4 \citep{Alessio2005}, T~Cha, LkH$\alpha$330, SR~21 and HD~135344B\footnote{HD135344B is also referred to as SAO~206462.} \citep{Brown2007} and many others, where the cavity size was inferred from radiative transfer modeling (see also Section \ref{sec:seds}). Follow-up imaging with millimeter interferometry with the SubMillimeter Array (SMA), CARMA and Plateau de Bure Interferometer (PdBI) confirmed the presence of inner dust cavities in several of these disks \citep[e.g.][]{Pietu2006,Brown2009,Hughes2009,Isella2010,Andrews2011}, although the recovered images had relatively low image fidelity due to the limited baselines and thus sparse uv-coverage of these interferometers and the derivation of the cavity size was primarily based on the location of the `null' in the visibility curve \citep{Hughes2007}. The maximum baselines of these interferometers reached a few 100 meters, resulting in a beam size of 0.3" at best, resolving only the largest inner cavities ($\gtrsim$30 au). Other observations in e.g. thermal infrared imaging also revealed inner dust clearing \citep{Geers2007} and deficits in the inner gas surface density through spectro-astrometry of near infrared CO rovibrational lines \citep[e.g.][]{Pontoppidan2008,vanderPlas2009,Brown2012a}. However, it was not until the start of operations of the ALMA telescope that the morphology of transition disks and their cavities in dust and gas could be fully revealed and quantified (see Section \ref{sec:mmobs}). 

\begin{figure}[!ht]
\centering
\includegraphics[width=0.8\textwidth]{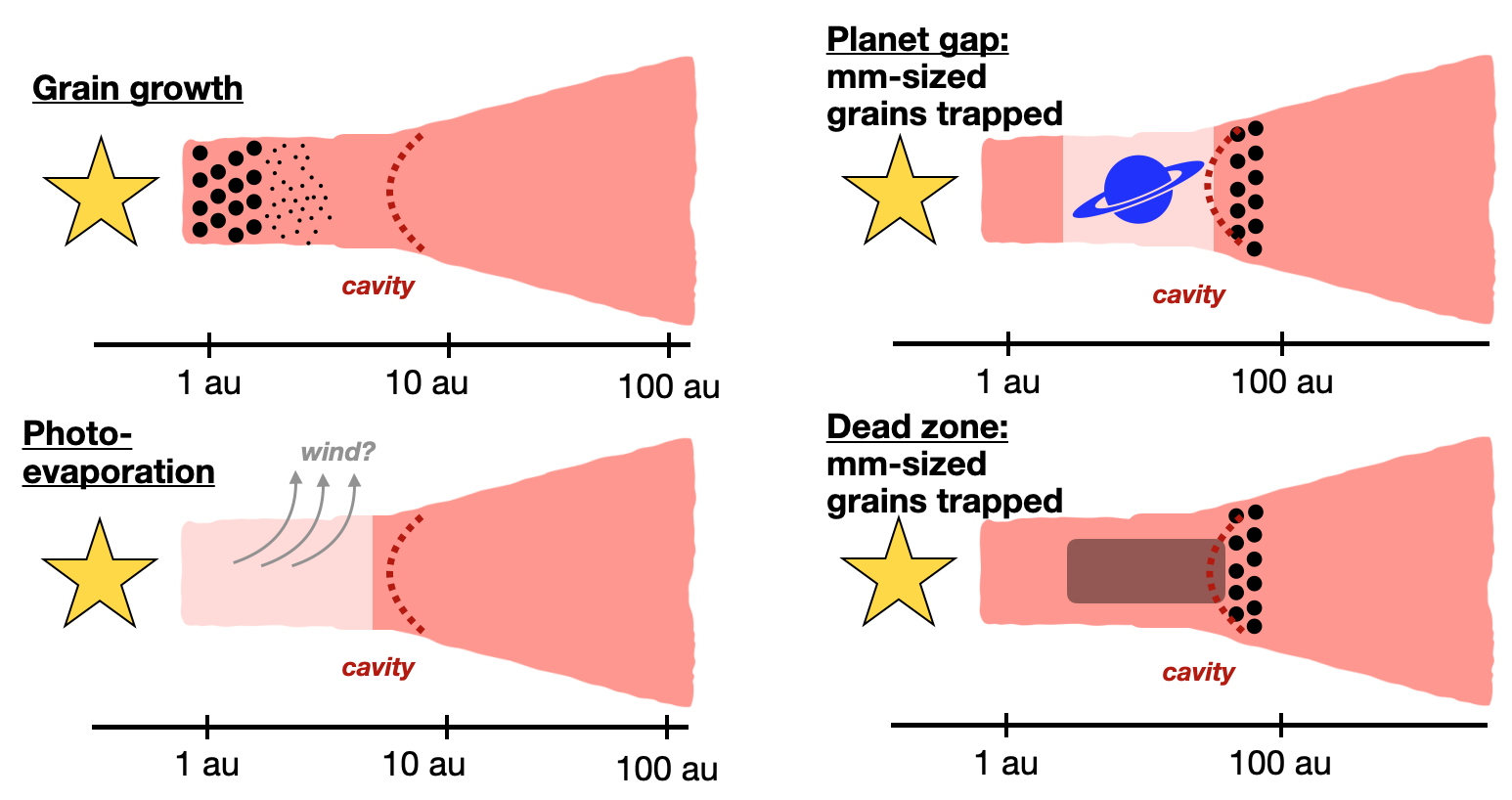}
\caption{The four main mechanisms proposed to explain transition disks: grain growth \citep{Dullemond2001}, photoevaporation \citep{Alexander2006b}, clearing by a companion \citep{Artymowicz1996} and dead zones \citep{Regaly2012}. The latter two processes are accompanied by dust trapping in a pressure bump \citep{Birnstiel2010}. More information in the text.}
\label{fig:mechanisms}
\end{figure}

Meanwhile, a large number of physical mechanisms had been proposed to explain the inner cavities in transition disks, including the dynamical clearing by companions \citep{Artymowicz1996,Zhu2011}, where the latter could be a stellar companion, i.e. the transition disk as circumbinary disk; photoevaporative clearing at the end of the lifetime of the disk  \citep{Alexander2006b,Owen2011}, or a combination thereof \citep{Alexander2009}, dead zones \citep{Regaly2012} or enhanced grain growth \citep{Dullemond2001}. The discovery of the 8 au stellar companion inside the 14 au cavity of CoKu Tau/4 \citep{Ireland2008} initially led to the suggestion that all transition disks might be circumbinary disks, and a binary companion was also discovered for HD142527 \citep{Lacour2016,Claudi2019}, but firm limits on companions in several transition disk cavities made this unlikely as a general explanation \citep{Kraus2011}. 

Several Jovian companion candidates have been identified and dismissed \citep[e.g.][]{Ligi2018,Currie2019,Wagner2019}, and it was not until the recent confirmation of the protoplanets in the PDS~70 and AB~Aur transition disks \citep{Keppler2018,Haffert2019,Currie2022} that planetary companions are considered a viable explanation for at least some transition disks, where many targets still lack firm enough limits due to the challenges in high contrast imaging in young disks \citep{Asensio2021,vanderMarel2021asymm,Currie2022PPVII}. Photoevaporative clearing was initially dismissed for the transition disks with high accretion rates and large ($>$20 au) inner cavities \citep{OwenClarke2012,Ercolano2017} leaving only the transition disks with smaller, SED-derived cavities as possible candidates, although recent efforts in photoevaporative modeling show that perhaps a wider range of cavities could be caused by photoevaporation as well \citep{Picogna2019,Ercolano2021}. Also other mechanisms have been proposed to explain cleared cavities (see Section \ref{sec:mech}). Overall, there may not be a single mechanism responsible for all transition disks and, also considering the range of definitions, it may be better not to consider transition disks as a single population, in particular when using their occurrence rates as a way to infer dissipation timescales. Specifically, transition disks should not be considered as a separate stage of disk evolution as long as it is not evident that all protoplanetary disks go through an observable 'transitional' phase. An empirical definition of a transition disk could lead to biases depending on the used selection criteria and this should be considered whenever transition disks are treated as a population.

`Complete' transition disk samples remain extremely challenging to define, both in SED studies and in millimeter imaging. In the latter, the spatial resolution limits the cavities that can be sufficiently resolved, in particular in the so-called snapshot surveys that target all Class II disks in nearby star forming regions which are always relatively low resolution for time efficiency \citep[for an overview, see e.g.][]{Manara2022}. In SED studies, transition disk candidates are usually selected with color criteria \citep[e.g.][]{Brown2007,Muzerolle2010,Cieza2010,Merin2010}, followed by radiative transfer modeling which is known to be highly degenerate and identification can be affected by e.g. extinction, irradiation effects and inclination \citep{Ercolano2009,Merin2010,vanderMarel2022}. Furthermore, cavity sizes from SED modeling are known to differ from those from millimeter imaging by a factor 2-3 \citep{Brown2009,vanderMarel2016-spitzer}, although the latter could also be the result of dust grain size segregation \citep{Rice2006,Zhu2011}. The definition of the word `transition disk' has become less definitive with the discovery of narrow-gapped disks such as e.g. HL Tau, HD163296 and several disks in the DSHARP survey \citep{HLTau2015,Isella2016,Andrews2018}. The inner cavities in transition disks could be considered as the most extreme cases of such gaps in terms of their width and location \citep{Andrews2020}. Considering their historic nature in both observational sense (deficit in the SED, resolved gap in pre-ALMA images) and possible underlying mechanisms (inside-out photoevaporative clearing or clearing by massive, Jovian planets), as well as their potential for well-resolved imaging due to the large cavity sizes compared to gapped disks, transition disks can still be considered as a separate group of interest in disk studies as long as biases are considered.

In this review I will discuss the properties of known transition disks within 200 pc and primarily focus on the disks with cavity sizes $\gtrsim15$ au to be as complete and unbiased as possible. I will discuss the limitations of this approach for our general understanding of transition disks and the steps that are necessary for a truly complete picture. Section \ref{sec:mmobs} describes the disk properties as derived from (primarily) ALMA millimeter imaging, essentially dismissing previous millimeter interferometry results due to the superb image quality and increase in both resolution and sensitivity of ALMA. Section \ref{sec:infrared} reviews the various insights from infrared observations of transition disks, although scattered light imaging is kept brief as this topic is largely discussed in the PPVII chapter by \citet{Benisty2022}. Section \ref{sec:seds} summarizes the insights from SED analysis and how these compare to the now known morphologies seen in millimeter dust continuum. In Section \ref{sec:analysis}, I will analyze a large number of properties of the known transition disks to infer trends and possible connections with the proposed mechanisms, followed by an outlook on future studies of transition disks with new and upcoming observational facilities.

\section{The millimeter perspective}
\label{sec:mmobs}

\subsection{Dust trapping}
\begin{figure}[!ht]
\centering
\includegraphics[width=0.9\textwidth]{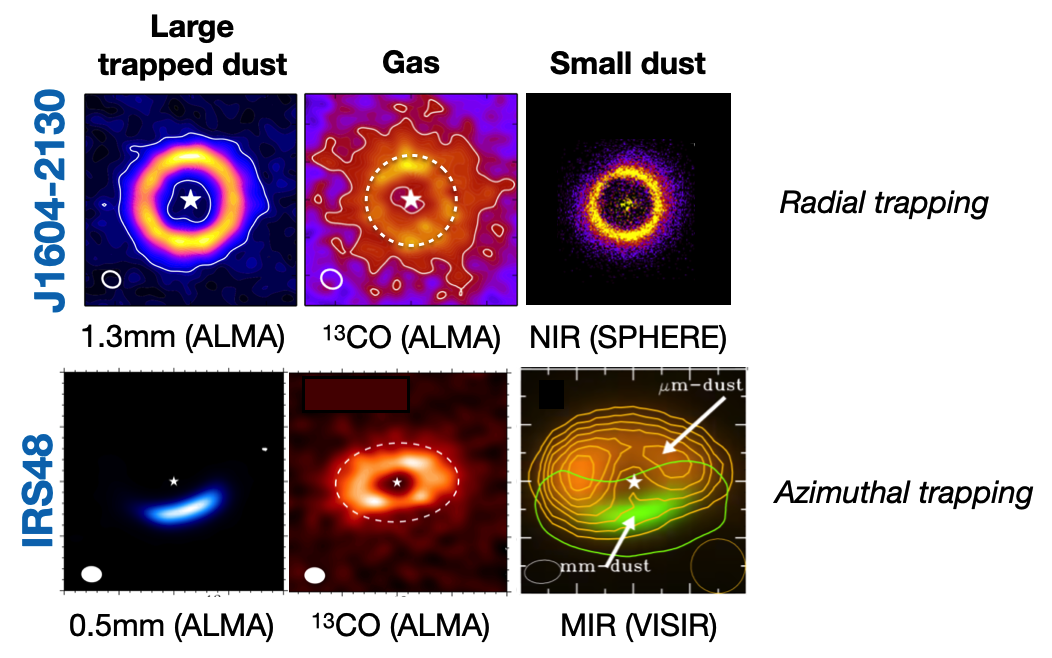}
\caption{Two examples of observational evidence for dust trapping, both radially (top) and azimuthally (bottom) for the J1604.3-2130 and Oph~IRS 48 transition disks based on millimeter continuum emission, integrated $^{13}$CO emission and infrared emission (originating from small dust grains). The white dashed ellipse indicates the radius of the millimeter dust.The segregation between the gas/small grains and millimeter dust provides evidence for the existence of trapping. Images combined from \citet{vanderMarel2013,vanderMarel2016-isot,Dong2017}.}
\label{fig:dusttrap}
\end{figure}

The millimeter images of transition disks have given evidence for the existence of dust trapping in protoplanetary disks, as a way to retain millimeter-sized dust particles in the outer part of the disk that would otherwise be transported inwards as the result of radial drift \citep{Weidenschilling1977}. Radial drift is the result of drag forces between the gas and dust grains, depending on their Stokes number (which is proportional to the grain size). As a smooth, axisymmetric protoplanetary disk naturally has a negative radial pressure gradient in the gas, millimeter-sized dust grains tend to drift inwards rapidly, essentially limiting the potential for further growth to planetesimals and planets \citep[e.g.][]{Brauer2008,Birnstiel2010}. The presence of local pressure maxima or pressure bumps in the outer disk have long been proposed as possible explanation for the presence of millimeter-sized dust grains in disks \citep[e.g.][]{BargeSommeria1995,Pinilla2012a}, as evidenced from strong millimeter fluxes and low spectral indices \citep[e.g.][]{Andrews2009,Ricci2010}. The pressure maxima act as so-called \emph{dust traps}, which limit the radial drift. Dust can be trapped in both the radial and azimuthal direction, leading to either ring-like or asymmetric millimeter-dust concentrations \citep[e.g.][]{Pinilla2012a,Birnstiel2013}, whereas the gas and small grains are still distributed throughout the disk with more moderate changes in the surface density. The segregation between millimeter-sized dust grains versus gas and small grains provides evidence for dust trapping (see Figure \ref{fig:dusttrap} and \citet{Rice2006}). 

Even before ALMA, transition disks had been proposed as promising candidates of disks with dust traps in the context of dust filtration between large and small grains to explain the presence of large inner cavities in combination with the high accretion rates \citep{Rice2006,Pinilla2012b,Zhu2012}. However, the highly asymmetric nature of the millimeter-dust continuum emission in the Oph~IRS 48 transition disk in contrast with the ring-like distribution of the gas as traced by $^{12}$CO 6--5 and the small grains as traced by VLT/VISIR 18 $\mu$m emission \citep{vanderMarel2013} provided the first observational evidence for the existence of dust traps. The smaller inner cavity seen in the $^{12}$CO 6--5 with a radius of $\sim$20 au with a dust trap at 60 au is consistent with a planet-gap model where the dust is trapped at the very outer edge of the gap \citep{Pinilla2012b}. Radial dust traps were identified soon after in e.g. J1604.3-2130 \citep{Zhang2014} which showed a narrow continuum dust ring and gas disk as traced with $^{12}$CO 3--2 with a much smaller inner cavity. Also scattered light images provided evidence for dust trapping, as the near infrared emission tracing the small grains in the disk showed much smaller inner cavities or no cavity at all (likely due to the coronagraphic mask) compared to the millimeter-dust emission \citep{Dong2012,Garufi2013}, which could be reproduced by dust evolution models with a dust trap at the edge of a planet-carved gap \citep{DeJuanOvelar2013}. Further developments in observations of gas and dust distributions in transition disks are described in Section \ref{sec:gasdust}.

Dust substructures, in particular gaps and rings, are now commonly seen in high-resolution ALMA continuum images of bright protoplanetary disks \citep[e.g. the DSHARP survey,][]{Andrews2018}, and usually directly interpreted as dust traps \citep{Dullemond2018}. Transition disks thus form a subset of protoplanetary disks with dust traps, and may be defined as a separate class of objects due to historic reasons, as they contain the first resolvable substructures on scales of $\gtrsim$30 au even with pre-ALMA interferometers. Smaller scale substructures of $\lesssim$5 au remain unresolved unless observed with the highest ALMA resolution \citep{Bae2022}. Furthermore, considering their potential origin of photoevaporative clearing, as well as circumbinary and eccentric disks (see also Section \ref{sec:clearing}), they remain interesting to study separately from gapped disks which cannot be explained by these mechanisms. Regardless of the scale and origin of the pressure bumps causing the dust substructures, it is clear that they play an important role in maintaining detectable millimeter dust in protoplanetary disks as radial drift leads to a decrease in observable dust mass \citep{Pinilla2020}. A recent disk demographic study suggests that only a subset of protoplanetary disks is able to maintain a high millimeter-dust mass due to the presence of dust traps, whereas the majority of disks decreases in dust mass due to radial drift in the absence of pressure bumps \citep{vanderMarelMulders2021}, although this remains up to debate in lack of high resolution ALMA observations for the majority of the fainter protoplanetary disks. 

\subsection{Millimeter-dust cavities}
The key feature of a transition disk in millimeter images is the cleared inner cavity in the dust continuum image. A `cleared' inner cavity implies that the inner part of disk is devoid of millimeter-size dust grains (except for the very inner disk, see below), with a depletion factor of several orders of magnitude based on radiative transfer calculations \citep{Andrews2011,vanderMarel2015-12co}. Cavities have been measured with a large range of sizes, ranging up to radii of $\sim$200 au in case of GG Tau (a circumbinary disk), although more typically sizes of $\sim$30-50 au are found. A lower limit to the cavity radius is difficult to assess as this is essentially limited by the spatial resolution of the observations, but cavity sizes as small as 5-10 au have been observed \citep[e.g.][]{Kurtovic2021}, as well as a 2 au gap in TW Hya \citep{Andrews2016}. Disks in nearby star forming regions within 200 pc have been imaged with ALMA with a spatial resolution of 0.2-0.6" in disk survey studies \citep[e.g.][]{Ansdell2016,Barenfeld2016,Pascucci2016,Long2018taurus,Cieza2019} whereas a subset of interesting (usually brighter) targets have been imaged at higher resolution of 0.05-0.1" as well, including many transition disks \citep[e.g.][]{Dong2017,Boehler2018,Facchini2019,Pinilla2019,Cieza2020,Mauco2021,vanderMarel2022}. For Lupus, the ALMA transition disk population identified from the low-resolution snapshot survey was studied separately \citep{vanderMarel2018}. Larger sample studies of transition disks usually consist of a range of spatial resolutions as they are constructed from ALMA archival data of multiple programs \citep{Pinilla2018tds,Francis2020}. A handful of transition disks have been identified in younger $<$1 Myr old systems as well, e.g. WL~17 in Ophiuchus with a cavity of 13 au \citep{Sheehan2017,Gulick2021} and 7 systems in Perseus with cavity radii from 30-100 au \citep{Sheehan2020}.

\begin{figure}[!ht]
\includegraphics[width=\textwidth]{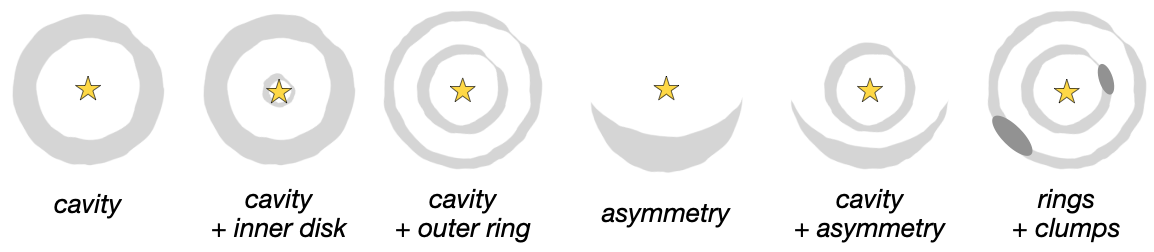}
\caption{Different types of millimeter substructures in transition disks.}
\label{fig:morphologies}
\end{figure}

The range in spatial resolution of the disk surveys limits the detectable transition disk cavities to a different minimum radius, depending on the survey, typically 0.25" or detectable cavity radii $\gtrsim$20 au in Lupus, Ophiuchus and Taurus \citep{Ansdell2016,Cieza2019,Long2019,Akeson2019}, although Taurus has not been studied uniformly with ALMA yet and some disks only have unresolved SMA observations \citep{Andrews2013}. Chamaeleon was imaged at 0.6" and only cavities $\gtrsim$50 au have been resolved \citep{Pascucci2016}, while Upper Sco was imaged at 0.35-0.55", revealing only two transition disks with a cavity of $\sim$85 au, although based on the resolution cavities $\gtrsim$25 au can be readily excluded in the remaining sample \citep{Barenfeld2016}. The detectable cavity radius can be a factor of 2-3 smaller than the angular resolution as the resolution measures the diameter, and modeling in the visibility plane can infer even smaller cavities. Additional transition disks have been identified in targeted studies at higher resolution down to cavity sizes of $\sim$10-15 au \citep[e.g.][]{Francis2020,Kurtovic2021,vanderMarel2022}. Overall, the sample of transition disks with cavities $\gtrsim$25 au within 200 pc is more or less complete based on these surveys, with possibly some additional unresolved transition disks in Chamaeleon and Taurus. In Section \ref{sec:comparison} I will review the known transition disks with large cavities from both SED and millimeter imaging studies and estimate the completeness of the known transition disks within 200 pc. 

With the higher sensitivity of ALMA it has become evident that transition disk cavities are not entirely cleared of millimeter-sized dust: centrally peaked emission cospatial with the star is regularly detected in cavities of fully resolved images of transition disks \citep[e.g.][]{Fukagawa2013,Pinilla2019,Pinilla2021,Francis2020,Facchini2020}. This emission is usually interpreted as thermal dust emission from the inner disk, although contributions of free-free emission cannot be excluded when the emission is unresolved, in particular at wavelengths $>$2 millimeter \citep{Macias2021}, see also Section \ref{sec:multiwave}. Inner dust disks have traditionally been inferred from near infrared excess in the SED, where the disk is also classified as `pre-transitional disk' \citep{Espaillat2007}. However, a lack of correlation between near infrared excess and the millimeter flux of the inner disk has cast doubt on the significance of this separate disk class \citep{Francis2020}, as further discussed in Section \ref{sec:innerdisksed}. The presence of inner dust rings also makes it harder to have a clearly distinct definition of a transition disk compared to so-called `ring disks' such as the disks in the DSHARP survey \citep{Andrews2018}, for example in the cases of HD~143006, DM~Tau and WSB~60 \citep{Perez2018dsharp,Kudo2018,Francis2020}. It has also become clear that not all transition disks are single dust rings (see Figure \ref{fig:morphologies}): some disks with inner cavities also contain one or more additional outer rings or asymmetries, which make them again similar to ring disks, for example HD~97048, HD~135344B, HD~169142, AA~Tau, GM~Aur and V4046~Sgr \citep[e.g.][]{Francis2020}. In this review paper these disks are not considered separately as the inner clearing mechanism can still be compared to that of other, `single-ring' transition disks. 

Cavity sizes in millimeter images can be derived using visibility modeling \citep[e.g.][]{Berger2007}, where the first `null' (the Real component crossing through zero) is a measure of the cavity size \citep{Hughes2007}. This was particularly relevant in pre-ALMA interferometric data due to the sparse uv-coverage of early millimeter arrays, where the images could often not directly reveal the presence of a cavity. Initially the emission profile was usually estimated using a radiative transfer model, fitting the SED and the millimeter data simultaneously using a power-law surface density profile with a steep dropoff at the cavity edge \citep[e.g.][]{Brown2009,Isella2010,Andrews2011,Zhang2014,vanderMarel2015-12co,Dong2017}. As higher resolution ALMA images started to reveal more detailed radial profiles (and asymmetric, see next section), the emission started to get fit directly with a parameterized intensity profile, usually a Gaussian which could be either symmetric or asymmetric around the peak radius \citep[e.g.][]{Pinilla2017,Pinilla2018tds} or even with multiple Gaussian components to fit multiple rings \citep[e.g.][]{Facchini2020}. The radial peak is used as equivalent to the cavity radius.

\subsection{Asymmetries}
\label{sec:asymmobs}
About 20\% of the observed transition disks in the sample study of \citet{Francis2020} show striking asymmetric morphologies in their millimeter dust rings, referred to as arcs, clumps, horseshoes or asymmetries \citep{Andrews2020,vanderMarel2021asymm}. These features are generally interpreted as azimuthal dust traps, as they are clearly distinct from the gas and small grain distributions. Asymmetries were tentatively detected pre-ALMA in e.g. HD142527 and AB~Aur \citep{Ohashi2008}, SR21 and HD135344B \citep{Brown2009}, Oph~IRS48 \citep{Brown2012b} and LkH$\alpha$330 \citep{Isella2013} but not always recognized as such due to low SNR, image fidelity and unknown stellar position. 

The first significant asymmetric dust rings were discovered in ALMA images of Oph~IRS48 and HD142527 \citep{vanderMarel2013,Casassus2013}, with azimuthal contrasts of a factor 130 and 30, respectively, on radial scales of 50-100 au. Also AB~Aur and LkH$\alpha$330 show a highly asymmetric dust ring \citep{Tang2017,vanderMarel2021asymm,Pinilla2022LkHa}. Shallower dust asymmetries with a contrast of a factor 1.5-2 were found in SR~21 and HD135344B \citep{Perez2014,Pinilla2014}, although in higher resolution observations it turned out that these actually consisted of multiple dust rings, some asymmetric, with higher contrasts \citep{vanderMarel2016-hd13,Cazzoletti2018,Muro-Arena2020}, as well as V1247~Ori, located at 400 pc \citep{Kraus2017}. These strong asymmetric features have been interpreted as dust trapping in either long-lived vortices as the result of the Rossby Wave Instability of a radial pressure bump \citep{BargeSommeria1995,Birnstiel2013,Zhu2014} or in gas horseshoes due to a pile-up of material in eccentric cavities in a circumbinary disk \citep{Ragusa2017,Miranda2017,Price2018}. The ultimate distinction between these two mechanisms is the $\alpha$-viscosity and the companion mass: vortices are only long-lived when $\alpha_t\lesssim10^{-4}$ but can be triggered by Jovian companions, whereas the gas horseshoe is also formed at moderate $\alpha_t\sim10^{-2}$ and requires a (sub)stellar companion with a mass ratio of at least 0.05 w.r.t. the host star \citep{vanderMarel2021asymm}. As current limits on both the $\alpha$-viscosity and companions are inconclusive in most asymmetric disks, the dominating mechanism remains unknown.

\begin{figure}[!ht]
\centering
\includegraphics[width=0.9\textwidth]{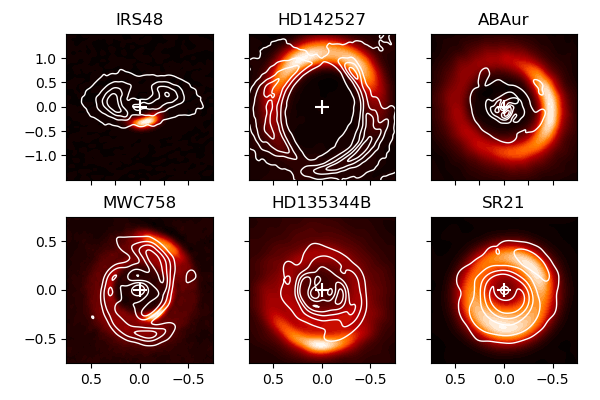}
\caption{Comparison between asymmetric dust features seen with ALMA and the spiral features seen in scattered light, data from \citep{vanderMarel2021asymm}. The top row shows large scale asymmetries which appear to be unrelated, whereas the bottom row show striking similarities in the morphologies, indicating that the asymmetry seen in millimeter wavelengths may be part of the spiral arm itself \citep{vanderMarel2021asymm}.}
\label{fig:asymm}
\end{figure}

Clumpy asymmetric features have been found in e.g. HD34282 \citep{vanderPlas2017}, HD143006 \citep{Andrews2018}, TW Hya \citep{Tsukagoshi2019} and MWC758 \citep{Boehler2018,Dong2018}, and spiral-like features in HD100453 \citep{Rosotti2020} and HD100546 \citep{Norfolk2022}, suggesting that some asymmetries could also be part of the spiral arms seen in scattered light images of the disk. Figure \ref{fig:asymm} shows a couple of examples of asymmetries and spiral arms seen in scattered light. Alternatively, it has been proposed that the underlying vortex triggers the spiral density wave itself \citep{Lovelace2014,vanderMarel2016-hd13,Cazzoletti2018}, although this has been debated by  simulations \citep{Huang2019}. More tentative asymmetric features have been identified in CQ~Tau \citep{Ubeira2019}, DM~Tau \citep{Hashimoto2021a} and ZZ~Tau IRS \citep{Hashimoto2021b}. 

Asymmetric features can be identified in both the image plane and the visibility plane: in the latter case, an asymmetric feature is apparent from non-zero emission in the imaginary visibility curve, assuming that the data is perfectly phase-centered on the center of the disk ring, properly deprojected and intrinsically flat. An incorrect phase center can result in the apparent presence of asymmetries, e.g. comparison of \citet{Pineda2014,Walsh2014}. Visibility modeling of an asymmetric disk can be performed with e.g. \texttt{galario} \citep{Tazzari2018}, providing a two-dimensional intensity profile rather than the default one-dimensional radial profile, as successfully demonstrated by e.g. \citep{Cazzoletti2018,vanderMarel2021asymm}. Care should be taken when fitting multiple components when they are not directly visible in the image: \citet{Pinilla2014} found a best fit for a ring+inner vortex model to ALMA visibility data of HD135344B, but re-imaging of the data revealed that the vortex was actually located outside of the ring \citep{vanderMarel2016-hd13}.

It has been proposed that the appearance of dust asymmetries is related to the local gas surface density, or more precise, the Stokes number of the emitting millimeter grains \citep{vanderMarel2021asymm}. This implies that asymmetric features may be more common at longer wavelengths, which can be tested with future facilities such as the ngVLA and ALMA Band 1 receivers. Another reason that more asymmetries may be detected at longer wavelengths is because the emission at (sub)millimeter wavelengths is generally optically thick \citep{Andrews2018ngvla}. Hints for asymmetric features at centimeter wavelengths in disks that are axisymmetric in millimeter wavelengths are seen in e.g. LkCa~15, SR~24S and J1604.3-2130 \citep{Isella2014,Norfolk2020}, but a higher image fidelity is required to confirm this phenomenon. 

Overall, the detection rate of asymmetric features in protoplanetary and transition disks remains a matter of sensitivity, spatial resolution and optical depth, and it is very likely that more disks turn out to be (somewhat) asymmetric in nature. Higher resolution observations generally lead to higher azimuthal contrast as dust rings remain radially unresolved at $>$0.1" resolution. Large scale asymmetries such as seen in Oph~IRS48 and HD142527 are unlikely to remain undetected in the disk surveys, so at least for that high-contrast asymmetry, the occurrence rate remains fairly low with less than 10\% of the known transition disks with large inner cavities ($>$20 au). Clearly, (almost) axisymmetric dust rings dominate the dust morphology population at large scales. Higher resolution observations of smaller disks are essential to confirm this trend. 

\subsection{Gas in cavities}
\label{sec:gasdust}
One of the key questions in transition disk studies is how much gas is still present inside the cavity, and what its surface density profile looks like. The gas is directly tracing the dynamics of the disk and thus may reveal the mechanism that is responsible for the clearing of the cavity. The presence of gas inside the cavity is evident from the observed high accretion rates in transition disks \citep[e.g.][]{Najita2007,Manara2014}, indicating that whichever clearing mechanism is responsible for the cavity, it allows material to be transported across the gap. Also kinematic analysis of CO rovibrational line emission indicates the presence of gas well inside the dust cavity of a small number of disks where the inner radius could be determined \citep[e.g.][and see Section \ref{sec:rovib}]{Pontoppidan2008,Brown2012a,Banzatti2018}. 

\begin{figure}[!ht]
\centering
\includegraphics[width=0.6\textwidth]{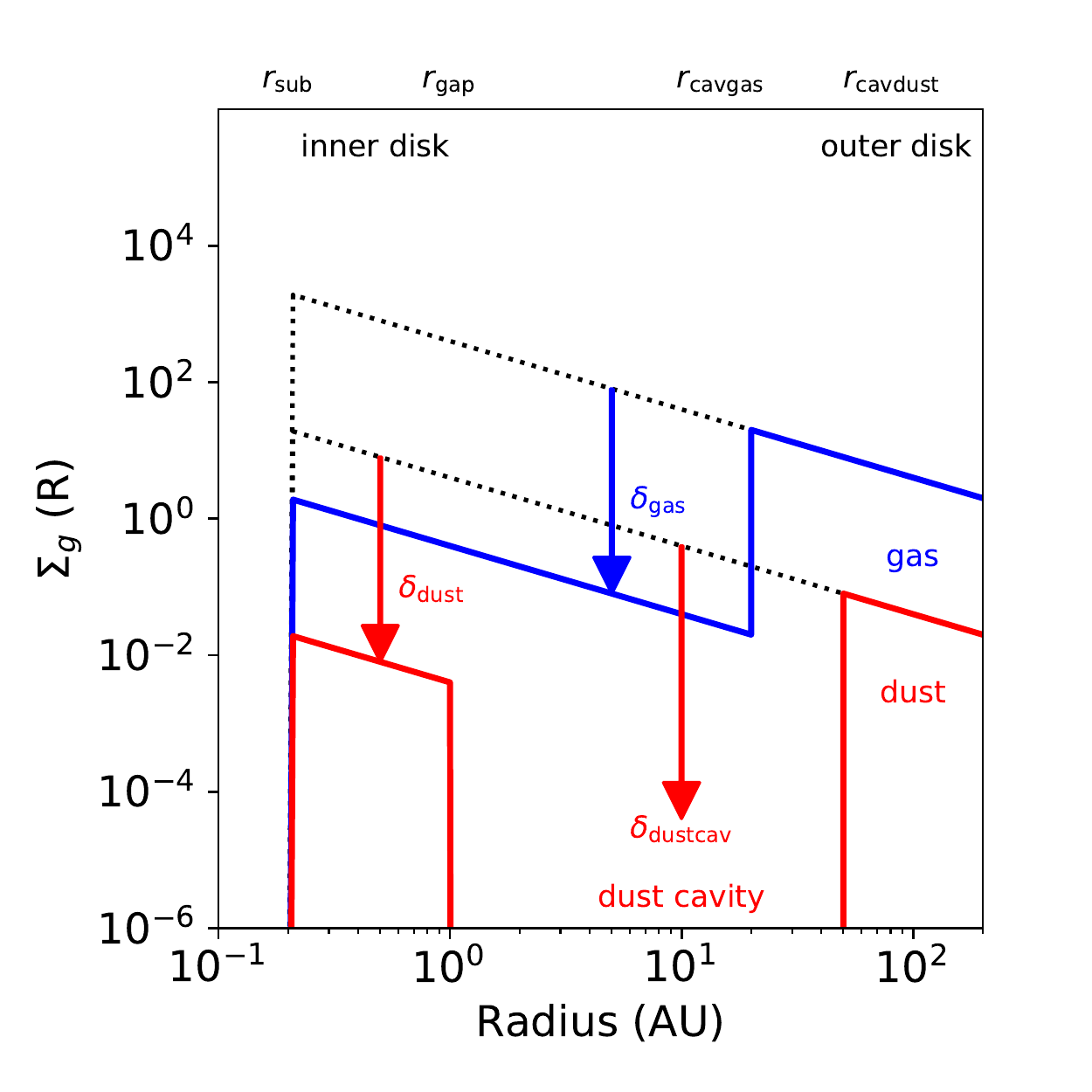}
\caption{Typical parametrized surface density profile of gas and dust in thermal-chemical modeling of transition disks. The gas cavity is generally smaller and less depleted than the dust cavity, whereas an inner dust disk is generally included to reproduce near infrared emission, if present. Figure based on \citet{vanderMarel2018}.}
\label{fig:surfdens}
\end{figure}

The relative gas surface density profile inside the dust cavity can be determined using ALMA CO isotopologue data, with sufficient resolution to resolve the cavity by at least one beam size. A combination of optically thick and thin emission, i.e. multiple isotopologues, is crucial for an accurate determination of the gas structure. Early ALMA observations revealed deep CO cavities in e.g. Oph~IRS48 \citep{vanderMarel2013} and J1604.3-2130 \citep{Zhang2014} which were a factor 3 smaller than the dust cavity radius, where the density dropped below the threshold of optically thick $^{12}$CO, at approximately 10$^{-2}$ g cm$^{-2}$. Considering self-shielding, it is unlikely that the CO itself (rather than H$_2$) is depleted in the inner part of the disk and the observed profiles thus represent an actual surface density drop \citep{Bruderer2013}. Other transition disks, such as HD142527, SR21 and HD135344B did not show a cavity in the $^{12}$CO emission \citep{Casassus2013,Perez2014,vanderMarel2015-12co}, implying that the cavity was either unresolved or the gas surface density remained above the optically thick threshold. 

Thermo-chemical modeling is required to quantify the gas surface density inside the dust cavity from the spatially resolved CO emission, for multiple reasons \citep{Bruderer2013}: the CO abundance is not constant w.r.t. H$_2$ throughout the disk due to freeze-out and photodissociation, although the latter can be diminished by self-shielding and dust shielding, in particular inside a dust cavity with an inner dust disk; the gas temperature is decoupled from the dust temperature, especially at the cavity edge and above the midplane, where optically thick lines originate; chemical reactions that lead to formation and destruction of CO vary throughout the disk depending on the local conditions. Dedicated modeling codes such as DALI and ProDiMo \citep{Bruderer2012,Woitke2009} can be used to calculate the expected emission cubes of targeted molecular lines based on a given density structure and radiation field. 

Spatially resolved images of CO isotopologue emission ($^{13}$CO and C$^{18}$O) revealed deep gaps inside the cavity of several transition disks, which was quantified through both thermo-chemical modeling \citep{Bruderer2014,vanderMarel2016-isot,vanderMarel2018} as well as simpler, parameterized radiative transfer modeling \citep{Perez2015,Boehler2017,Boehler2018}. The model fitting procedure assumed a  power-law surface density profile with drops in surface density at different radii for the gas and the dust, respectively (see Figure \ref{fig:surfdens}). It was found that the gas surface density dropped by several orders of magnitude inside the cavity, which points towards the presence of super-Jovian planets \citep{Fung2016,vanderMarel2016-isot}. However, this parametrized approach is physically unrealistic as the inward extrapolated gas surface density at 1 au was well below the expected value from the accretion rate, and because the gas surface density is not expected to drop steeply as the result of planet-gap clearing \citep{Pinilla2012b}. 

Later studies revealed that the three main CO isotopologues $^{12}$CO, $^{13}$CO and C$^{18}$O show increasing gap sizes, indicating a gradual dropoff where each isotopologue becomes optically thick at a different radius \citep[seen e.g. in HD~169142,][]{Fedele2017}, which was fit with a gradual slope for J1604.3-2130 and CQ Tau \citep{Dong2017,Ubeira2019}. Generally, the $^{13}$CO cavity radii are about a factor 2 smaller than the dust cavity radii \citep{vanderMarel2021asymm}. Planet-disk interaction models were implemented directly in the comparison with the observed CO profiles for PDS70 \citep{Muley2019} and in generic model predictions \citep{Facchini2018gaps}. The latter study showed that stronger central $^{13}$CO emission is expected as the result of planet clearing, and that this cannot only be explained by lack of spatial resolution, as previously claimed \citep{vanderMarel2018}. More recent observations fully resolve the cavity even in the CO emission, revealing some central emission in both $^{12}$CO and HCO$^+$ \citep[e.g.][]{Mayama2018,Long2018,Tsukagoshi2019} which allows further exploration of the link with the high accretion rate, such as recently done for DM~Tau \citep{Hashimoto2021a,Francis2022}. CO observations of multiple line transitions are required to fully unravel the gas temperature structure inside the cavity for a proper interpretation of the gas surface density profile \citep{Leemker2022}. Also, for TW Hya, the gas surface density profile inside the cavity could be determined from the direct measurement of pressure broadening in CO line wings \citep{Yoshida2022}.

Another aspect of gas studies in transition disks is the study of the kinematics and the brightness temperature of $^{12}$CO, which have revealed spiral signatures comparable to those seen in scattered light \citep[e.g.][]{Casassus2021,Wolfer2021,Wolfer2022}. Also warps are commonly seen in the velocity maps of transition disks \citep{Casassus2013,Boehler2017,Loomis2017,Mayama2018,vanderMarel2021asymm,Wolfer2022} which are generally interpreted as misalignment between the inner and outer disk as the result of a companion \citep[e.g.][]{Rosenfeld2014,Facchini2018,Zhu2019}, also revealed by azimuthally moving shadows \citep{Pinilla2018j1604,Nealon2020}. Other kinematic signatures such as kinks, meridional flows and circumplanetary disks \citep{Teague2018,Pinte2018,Pinte2019,Perez2019,Izquierdo2021} are harder to detect in transition disk cavities due to the drop in CO emission, and such signatures may only be found in gapped disks where the CO emission is less depleted.

\subsection{Multi-wavelength data} 
\label{sec:multiwave}
Multi-wavelength studies at millimeter and centimeter wavelengths (the latter with e.g. VLA or ATCA) are considered a powerful tool to learn more about the grain sizes in the disk, through the spectral index $\alpha_{mm}$ where values between 2 and 3 are generally interpreted as evidence for grain growth (in contrast with the $\alpha_{mm}$ ISM value of 3.7), although high optical depth could also result in lower $\alpha_{mm}$ \citep{Testi2014}. Disk-integrated spectral indices of a sample of transition disks were found to be higher than for regular disks, which can be understood as transition disks lack the inner emission which is often very optically thick \citep{Pinilla2014}. Spatially resolved multi-wavelength analysis of transition disk dust rings shows that the dust ring is narrower, and the peak is often slightly further out \citep{Pinilla2014,Pinilla2017-sr24s,Cazzoletti2018,Macias2019,Huang2020,Mauco2021}, although this can not always be confirmed, in particular with low resolution ATCA images  \citep{Norfolk2020}, and in a few cases the dust ring is even closer in at longer wavelengths \citep{Macias2018,Pinilla2021}, although this could also be caused by contributions from the central emission not included in the fit. Generally these data are consistent with predictions from dust evolution models \citep{Pinilla2018tds}.

Azimuthal asymmetries are in practice easier to study and compare in multi-wavelength observations than axisymmetric disks, as variations can be measured at larger scales and thus lower spatial resolution. Dust evolution models predict that the azimuthal extent decreases with wavelength in azimuthal dust traps \citep{Birnstiel2013}, which was indeed measured in Oph~IRS 48, HD142527 and HD135344B \citep{vanderMarel2015vla,Casassus2015,Cazzoletti2018}, while this remains inconclusive for SR21, MWC758 and AB~Aur \citep{Pinilla2014,Muro-Arena2020,Marino2015,Fuente2017}. 

At longer wavelengths, free-free emission originating from ionized gas close to the star \citep{Snell1986} will start to contribute to the total flux as well. This emission generally starts to dominate over the thermal dust emission below 30 GHz, and $\alpha_{mm}$ generally ranges from 0 to 1 in this regime \citep{Rodmann2006}. Several transition disks have been observed in the 10-30 GHz range revealing that the spectral index indeed flattens, consistent with free-free emission \citep{Ubach2012,Zapata2016}, although the emission can be highly variable \citep{Ubach2017}. In two cases, AB~Aur and GM~Aur, the emission is sufficiently resolved to distinguish a jet morphology rather than a disk \citep{Rodriguez2014,Macias2016}, but in most cases the origin of the emission (thermal or free-free) remains unknown by lack of resolution. In some cases, in particular due the azimuthal asymmetries, the central emission can be spatially distinguished in the centimeter emission \citep{vanderMarel2015vla,Marino2015,Casassus2015}, and also ALMA images generally resolve the central emission inside dust cavities \citep{Francis2020}, which opens up new possibilities for the study of the role of free-free emission in transition disks.

\section{The infrared perspective}
\label{sec:infrared}
\subsection{Infrared imaging}
Transition disks are a prime target of interest of near infrared imaging. At these wavelengths, observations probe the light scattered off by the surface layer of the disk and is thus tracing an intrinsically different regime than the millimeter images. Since the stellar emission dominates in the near-IR, observations are carried out with adaptive optics-aided instruments such as VLT/NACO, VLT/SPHERE, Gemini/GPI and Subaru/HICIAO and differential imaging techniques such as the Polarization Differential Imaging (PDI) that probes the polarized fraction of the scattered light.

Transition disks have been included in targeted surveys such as SEEDS \citep{Uyama2017}, Gemini-LIGHTS \citep{Rich2022}, DARTTS-S \citep{Avenhaus2018,Garufi2020}, and other SPHERE surveys \citep{Garufi2017,Garufi2018}. The reason for the high occurrence of transition disks in these surveys is two-fold: first, transition disks have always been chosen as key targets to look for wide-orbit companions due to their inner cavities and second, transition disks tend to be brighter in infrared wavelengths due to the irradiated cavity and due to being more common around the more luminous Herbig stars than T Tauri stars \citep{vanderMarelMulders2021} which makes them more suitable for AO correction with an optical wavefront sensor, although  almost the entire parameter space of stellar mass and age has been covered in near infrared imaging \citep[see Fig. 3 in][]{Benisty2022}.

For a full overview and discussion of the techniques and observed morphologies of near infrared imaging I refer to the PPVII chapter by Benisty \citep{Benisty2022}, I will merely highlight a few aspects here. Companions will be discussed in section \ref{sec:companions}. The near infrared images have revealed a wealth of substructures, such as spiral arms, gaps, rings, cavities and broad and narrow shadows as the result of misalignments, and in a few cases, companions and companion candidates, as discussed in the review chapter. The emission is usually located closer in than the millimeter dust ring, consistent with predictions from dust trapping models \citep{DeJuanOvelar2013}. Most substructures are linked to the presence of companions in the disk, even when those companions often remain undetected \citep{Benisty2022}. In particular spiral arms have received a lot of attention in connection with deriving companion properties in transition disk cavities \citep{Dong2015spirals,Fung2015,Zhu2015} but also wide outer companions. Alternatively, they have been connected with the morphologies of millimeter dust asymmetries in transition disks \citep[see discussion in e.g.][]{Garufi2018,vanderMarel2021asymm}, but are not uniquely found in cavities alone. A possible explanation for their high occurrence in transition disks was proposed by \citep{vanderMarel2021asymm} who noticed that spirals were predominantly found in disks with larger gaps and more luminous stars, suggesting that spiral arms are more easily detected for higher mass planets and warmer disks. 

Mid-infrared thermal emission can also be used to trace small grains in disks. Spatially resolving disks in infrared wavelengths requires ground-based 10m-class telescopes, and thus exceptional weather conditions to observe in the infrared, which means that such images are relatively rare. Oph~IRS48 was imaged with VLT/VISIR, where a dust ring was spatially resolved at 19 $\mu$m,  \citep{Geers2007}. Images with Subaru/COMICS revealed additional east-west asymmetries in the Oph~IRS48 disk \citep{Honda2018}. In sample studies of Herbig stars it was shown that mid infrared images of Group I disks are more often extended, i.e. warmer by direct illumination due to the presence of an inner hole, confirming that Group I Herbigs are transition disks with inner cavities whereas Group II disks generally do not have a cavity \citep{Maaskant2013,Honda2015}, indicating that there is no evolutionary transition from Group I to II disks  \citep[][and see Section \ref{sec:herbigs}]{Garufi2017}.

\subsection{Near infrared interferometry}
Transition disks with near infrared excess have been targeted in near infrared interferometry studies in H and K band to gain more insights on their inner dust disks. Early infrared interferometry with e.g. VLTI/AMBER, VLTI/MIDI and the Keck Interferometer revealed gap-like structures as well as dust composition in e.g. HD100546, T Cha and V1247 Ori, as well as HD179218 and HD139614 which have so far not been resolved with millimeter interferometry \citep{Benisty2010,Mulders2013,Olofsson2013,Kraus2013,Matter2016,Kluska2018}. A sample study with MIDI \citep{Menu2015} showed evidence for dust gaps in a large number of transition disk candidates (Group I disks, see Section \ref{sec:herbigs}).

More recently, larger samples of Herbig stars were studied with VLTI/PIONIER and VLTI/GRAVITY including several transition disks down to scales of a few tenths of an AU \citep{Lazareff2017,Perraut2019,Kluska2020}. Image reconstruction showed that inner disks contain both axisymmetric and non-axisymmetric structures, and that the inner rim location scales with the stellar luminosity. A targeted transition disk study with GRAVITY (including some of the brighter T Tauri stars) confirmed the misalignment of several of the inner dust disks with respect to the outer disk as traced by ALMA line observations, where the shadows seen in scattered light could lift the degeneracy in inclination angle \citep{Bohn2022}. 

Whereas near infrared interferometry is only sensitive to the inner rim of the inner disk, the MATISSE instrument which operates between L and N band, will be able to trace the dust emission out to larger radii of several au, which will provide further insights in the inner dust disk structure. 

\subsection{Infrared line emission}
\label{sec:rovib}
Rovibrational line emission in the infrared traces the hot and warm gas at radii within a few au of the host star, which is a valuable tracer of the physical processes inside transition disk cavities even when spatially unresolved. Kinematic analysis of rovibrational CO line profiles indicate the presence of gas in the inner disk \citep{Najita2008}, although many line profiles are single-peaked and thus more consistent with disk winds rather than Keplerian disk rotation \citep{Salyk2007,Salyk2009,Najita2009,Bast2011}. For some of the transition disks, the radial extent of the CO could be determined using spectro-astrometry with VLT/CRIRES, for example in SR~21, HD135344B, Oph~IRS48, HD100546 and HD139614 \citep{Pontoppidan2008,Brown2012a,Brittain2009,Brittain2013,Carmona2014,Carmona2017}, showing that the gas cavity radius is well inside the dust cavity radius. A broader sample study shows that transition disks rarely have a hot CO component in rovibrational line emission in comparison with full disks, indicating a removal of inner gas \citep{Banzatti2015}. Furthermore, \citet{Banzatti2018} find a link between the amount of refractory elements, near infrared excess and inner radius of the CO in a sample of Herbig transition disks, which is interpreted as full depletion of the gas inner disk in the presence of a planet and a dichotomy in the amount of near infrared excess. However, such a dichotomy was not seen in the millimeter dust mass of the inner disks in these transition disks \citep{Francis2020}, challenging this  interpretation.

Mid infrared spectra with e.g. \emph{Spitzer} generally lack strong molecular line emission in transition disks that is seen in full disks around T Tauri stars \citep{Najita2010,Pontoppidan2010}. In particular, the H$_2$O emission is found to be anti-correlated with disk dust size for a large sample of disks, which can be understood as icy pebbles are unable to drift inwards in the presence of dust traps, such as those found in transition disks \citep{Banzatti2020,Kalyaan2021}. Also for TW~Hya it has been suggested that the presence of dust traps is the main reason for the low elemental abundances of carbon and oxygen, considering infrared observations of CO and H$_2$O \citep{Bosman2019}. Dust transport and trapping may play a crucial role in setting the inner disk composition as traced by infrared emission, and transition disks are clear laboratories for quantifying these mechanisms, in particular in the coming years with the arrival of infrared spectra of inner disks with the \emph{James Webb Space Telescope}.

\section{The SED perspective}
\label{sec:seds}
\subsection{SED of a transition disk} 
\label{sec:sedtd}
As transition disks were originally identified through a deficit in their SED and millimeter interferometric imaging was not available until the 2000-2010s, many early efforts in transition disk studies were entirely based on the radiative transfer modeling of SEDs to derive properties of their dust structure based on spatially unresolved photometry \citep[see also the historic overview by][]{Espaillat2014}. \emph{Spitzer} played a crucial role here, providing mid and far infrared photometry of thousands of YSOs in nearby star forming regions, and targeted low-res spectroscopy with the IRS instrument in the 5-35 $\mu$m regime, where the deficit is most easily recognized. Unfortunately \emph{Spitzer} IRS spectra have only been taken for a limited number of targets so a uniform study is not possible \footnote{The reduced \emph{Spitzer} IRS spectra are available online \citep{Lebouteiller2011}.}. Transition disks were historically identified by comparison of the SED with the so-called `median' SED of full disks in Taurus, the best studied star forming region at the time. SEDs of transition disks usually show no excess emission in the near and mid infrared (1-5 and 5-20 $\mu$m) above the stellar photosphere, and excess at the level of the median Taurus disk at wavelengths $>$20 $\mu$m \citep{Calvet2005}, indicating a lack of hot and warm dust close to the host star. As 80\% of the emission at 10 $\mu$m originates from within 1 au \citep{Alessio2006}, the mid infrared spectrum is particularly sensitive to small cavities, and several transition disks that were imaged with millimeter interferometry turned out to have much larger cavities than originally thought based on the SED \citep[e.g.][]{Brown2009,Calvet2002,Andrews2011}. Some SEDs also show excess at 1-5 $\mu$m, which are sometimes called pre-transition disks in the literature \citep{Espaillat2007}, but there are some issues with this subclassification that will be discussed in the next section.

SEDs can be modelled using radiative transfer codes such as RADMC or MCFOST, assuming an inwardly truncated optically thick disk, where the cavity edge/wall of the outer disk is frontally illuminated by the star, which dominates most of the emission at 20-30 $\mu$m \citep{Brown2007}. The SED can constrain the vertical and radial structure of the small dust grains, but due to the illumination and shadowing of the inner disk, if present, the cavity size parameters are degenerate \citep{Dullemond2003}. The SED has also been used to constrain the amount of settling of large grains in the disk \citep{Dullemond2004,Cieza2007,Grant2018}. The total dust mass is only reliable when millimeter fluxes are included. Other properties, such as inclination and extinction, may also affect the outcome of the SED modeling \citep{Merin2010}. Finally, the 10 $\mu$m silicate feature is sometimes used as indicator of small, optically thin dust inside the cavity \citep{Espaillat2007}. With the advance of spatially resolved images of dust and gas, SEDs are often fit together with the images for better constraints on the structure \citep[e.g.][]{Andrews2011,vanderMarel2016-isot}.

Transition disk candidates have been identified from SED photometry using various color criteria indicating steep slopes in the infrared part of the SED \citep[e.g.][]{Brown2007,Furlan2009,Cieza2010,Merin2010}, that were easy to calculate for the thousands of YSOs in \emph{Spitzer} photometric catalogs of nearby star forming regions \citep[e.g.][]{Merin2008,Luhman2008,SiciliaAguilar2009}. Dedicated surveys of transition disk candidates using color criteria and subsequent SED analysis in the \emph{Spitzer} c2d and Gould Belt catalogs resulted in many new candidates \citep{Cieza2010,Cieza2012,Romero2012,vanderMarel2016-spitzer}, as well as contributions from \emph{Herschel} photometry \citep{Ribas2013,Bustamante2015,Rebollido2015}, where \citet{vanderMarel2016-spitzer} is still the most complete catalog of transition disk candidates to date, including estimates of cavity sizes and comparison of various criteria. 

\subsection{Herbig disks}
\label{sec:herbigs}
The study of Herbig disks and their SEDs has evolved almost completely separate from the aforementioned SED studies around the later type T Tauri stars which are much more common in the IMF. Herbigs are typically defined as B, A and F type young stars and have been identified both isolated and in clusters across their sky \citep[for a full overview, see e.g.][]{Vioque2018}.  The NIR images and the infrared color ($F_{30{\mu}m}/F_{13{\mu}m}$) divide the Herbigs into two groups: extended, bright disks with spiral arms and high $F_{30{\mu}m}/F_{13{\mu}m}$ value are called Group I, whereas faint, flat and small disks with low $F_{30{\mu}m}/F_{13{\mu}m}$ value are called Group II \citep{Meeus2001,Garufi2017}. Historically the two groups were considered to follow an evolutionary sequence from flared (Group I) to flattened (Group II) \citep{Dullemond2004}. Millimeter images of Group I disks all show large inner dust cavities and massive outer dust disks \citep[e.g.][]{Garufi2018}, whereas Group II disks are faint and appear to be compact in millimeter emission \citep{Stapper2022}. As the distinction appears to be about the presence or absence of an inner cavity, the groups are recently  considered as following two separate evolutionary pathways \citep{Maaskant2013,Garufi2018}. It is evident that almost all Group I disks are essentially transition disks, and they can be considered as the higher stellar mass ($M_*>1.5 M_{\odot}$) equivalent of T Tauri stars with transition disks. 

\subsection{Near infrared excess}
\label{sec:innerdisksed}
Near infrared excess in SEDs of transition disks is traditionally associated with the presence of an inner dust disk with dust down to the sublimation radius and a dust `gap', where a lack of such excess indicates a fully cleared cavity. A gapped disk is also called a pre-transition disk, as it was assumed that the inner dust disk would disappear over time as the companion clears out a wider gap and this is an evolutionary sequence \citep{Espaillat2007}. However, there are several issues with this interpretation: 1) The near infrared excess is primarily sensitive to the optically thick inner wall and thus not sensitive to the \emph{amount} of dust in the inner disk; 2) There is no correlation between the presence of the inner dust disk at millimeter wavelengths and the near infrared excess \citep{Francis2020}; 3) Some transition disks without near infrared excess have an inner dust ring that is just outside the sublimation radius, e.g. DM~Tau and TW~Hya while the companion responsible for the cavity is likely located further out \citep{Andrews2016,Hashimoto2021a,Francis2020}; 4) There is no clear difference between accretion rates of transition disks and pre-transition disks \citep{Espaillat2014}; 
5) The near infrared excess depends on the inclination of the inner disk, such as demonstrated for e.g. RY~Lup \citep{vanderMarel2018};  6) The near infrared excess can be variable with time, as seen in a comparison of \emph{Spitzer} and \emph{WISE} photometry in J1604.3-2130 \citep{Zhang2014}. Therefore, in this review the term pre-transition disk is not further used.

Some transition disks show `seesaw variability' in multi-epoch infrared observations, i.e. a decrease in near infrared flux is accompanied with an increase at longer wavelengths \citep{Muzerolle2009,Espaillat2011,Flaherty2012}, where the flux changes by 10-50\% over timescales of a few years. This variability is interpreted as a change in the structure of the inner disk, e.g. a change of the height of the inner disk with time, although the underlying physical mechanism remains unclear \citep{Espaillat2014}. Unfortunately this type of study was only possible by combining the \emph{Spitzer} IRS spectra taken in multiple cycles, which has only been done for a small number of targets. When comparing the seesaw sources from the variability study by \citet{Espaillat2011} with spatially resolved ALMA observations \citep{Francis2020}, there is no consistent picture: both seesaw and non-seesaw disks have a variety of inner disk dust masses or turn out not to have inner cavities at all \citep[e.g. WW~Cha and CR~Cha,][]{Ribas2013,Kanagawa2021,Kim2020}. Understanding this variability behaviour thus requires significantly more data and targets. 

Another type of variability is seen in optical emission of so-called `dippers' which can dim by a few tens of percent on timescales of approximately one day \citep[e.g.][]{Alencar2010,Cody2014}. This behaviour is thought to be caused by occultation of circumstellar dust, in particular edge-on disks \citep{Bouvier1999}. However, ALMA observations have revealed that dipper stars have a wide range of inclinations, even completely face-on disks such as J1604.3-2130 \citep{Ansdell2016,SiciliaAguilar2020}, suggesting that misalignments between inner and outer disks or warps may be common \citep{Ansdell2020}, which supported by numerous other studies of transition disks \citep[e.g.][]{Francis2020,Bohn2022,Wolfer2022}. Due to the heterogeneous studies of transition disks, it is not possible to assess the commonality of warps statistically. Dippers are not restricted to transition disks only and thus could be related to a range of phenomena \citep{Ansdell2020}. 

\subsection{Evolved disks}
Some SEDs of protoplanetary disks show decreased emission at all wavelengths rather than a decrease only up to 20 $\mu$m like in transition disks. Such disks are called `anemic' \citep{Lada2006}, `homologously depleted' \citep{Currie2011}, `evolved' \citep{Hernandez2007} or `weak excess transition' \citep{Muzerolle2010} and are sometimes grouped together with transition disks as evolved disks when calculating disk fractions and lifetimes \citep[e.g.][]{Muzerolle2010,Ribas2015} whereas they may represent completely different evolutionary processes in the disk. Such disks often also classify as Class III disks in the Lada classification which are known to be often mistaken in cluster membership in pre-Gaia studies \citep{Oliveira2009} and are only sparsely studied with ALMA \citep{Hardy2015,Lovell2021,Michel2021}. A systematic reassessment of these evolved Class III disks in both membership and dust mass would be highly beneficial for our general understanding of disk evolution, as well as their comparison with transition disks to determine if disks evolve along separate evolutionary pathways.

\subsection{Comparison between SEDs and millimeter imaging}
\label{sec:comparison}
A catalog of SED-derived transition disk candidates in \citet{vanderMarel2016-spitzer} is constructed using a combination of photometric color criteria with \emph{Spitzer} data and additional candidates from the literature. The main color criteria were based on the \emph{Spitzer} photometry: 0$<$[3.6]-[8.0]$<$1.8 and 3.2$<$[8.0]-[24.0]$<$5.3 \citep{Merin2010}. It thus provides the most complete sample of transition disk candidates. In order to investigate the correlation between SED selection and actually resolved millimeter cavities (and thus the completeness of the current known transition disks with cavities in the millimeter continuum) I have constructed a list of all transition disk candidates from this work within 200 pc with an estimated cavity size of 15 au or larger in Table \ref{tbl:sedcandidates}. For the candidates with cavity sizes $<$15 au such an assessment cannot be done as most disks have not been resolved at sufficient spatial resolution. A handful of targets were excluded because they are actually Class III, the \emph{Spitzer} photometry is contaminated \citep{Ribas2013} or because they turn out not to be cluster members according to Gaia DR2 catalogs \citep{Galli2019}. In addition, I have included all Group I Herbig disks within 200 pc that were not already included from \citet{vanBoekel2005} as the study by \citet{vanderMarel2016-spitzer} was not complete w.r.t. Herbig stars. 

\begin{figure}[!ht]
\centering
\includegraphics[width=\textwidth]{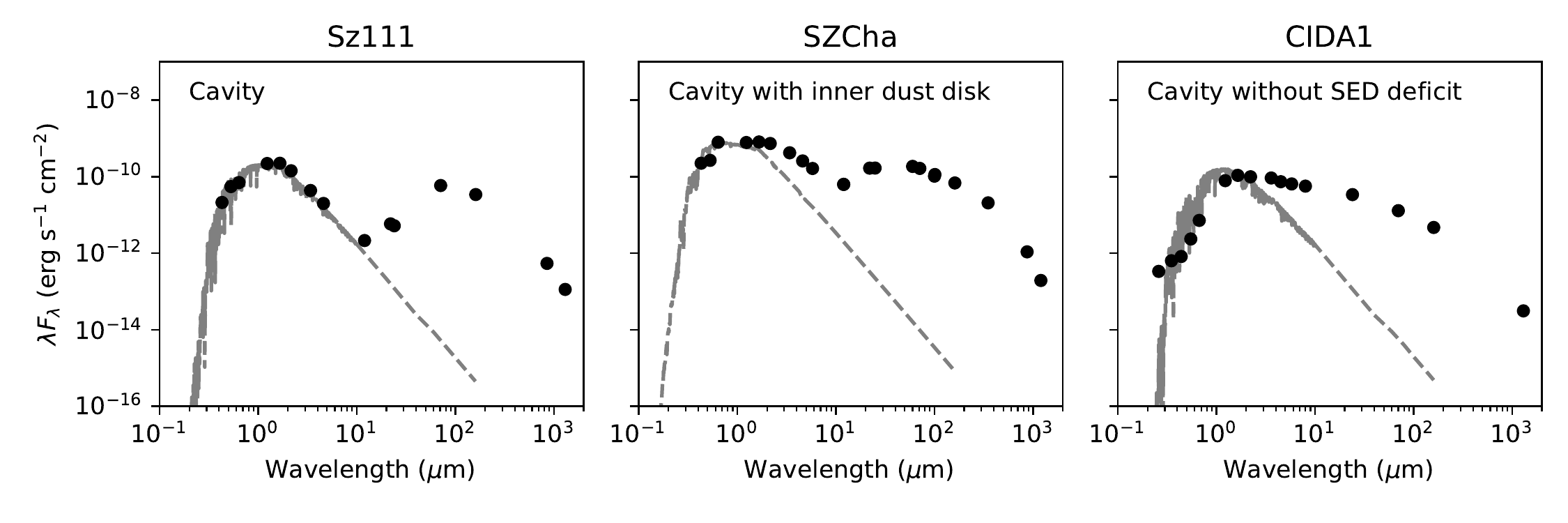}
\caption{Three examples of SEDs of transition disks with large inner cavities. A stellar spectrum is shown in dashed grey to show the amount of  infrared excess.}
\label{fig:seds}
\end{figure}

Table \ref{tbl:sedcandidates} contains 53 transition disk candidates, for which 34 have confirmed millimeter cavities with sizes ranging from 11 to 185 au (32 disks have a confirmed cavity size $>$15 au). For 13 disks there is no (sufficiently resolved) millimeter data available to make an assessment about their cavity size, for 6 disks a cavity $>$20 au can be excluded. In one case no dust emission was detected at all (V410X-ray6), one disk was very compact (CIDA~7), one disk was highly inclined (MY~Lup) and the other 3 disks are located in Ophiuchus where the SED suffers from high extinction which affects the SED modeling \citep{Merin2010}.  Excluding the 13 disks with insufficient data, this implies that 34/40 = 85\% of the SED selected transition disk candidates with $\gtrsim$15 au cavity indeed show a millimeter cavity. This could be as high as 47/53 = 89\% if the 13 disks without sufficient data will show a cavity as well, and as high as 34/36 = 94\% when highly inclined and highly extincted sources are excluded.

Second, Table \ref{tbl:newtds} contains a list of all published transition disks with cavities $>$15 au discovered from millimeter imaging that were not recognized as candidates by \citet{vanderMarel2016-spitzer} and references therein. This is a total of 22 transition disks. Perhaps surprisingly, the famous PDS~70 and AA~Tau transition disks are part of this list as well. Inspection of their SEDs and their classification in \emph{Spitzer} papers indicates that the majority of these transition disks do not show a clear deficit in their SED, which is the reason why they were not identified from color criteria. A handful of targets were not (or incompletely) measured by \emph{Spitzer} so color criteria could not have been applied. The reason for a lack of deficit could be caused by for example a misaligned optically thick inner disk such as seen in RY~Lup \citep{vanderMarel2018} and a handful of others \citep{vanderMarel2022}, but it is also possible that the $\mu$m-sized grains are not as depleted as the mm-sized grains in the cavity to decrease the infrared excess, which is often seen in the comparison between scattered light and millimeter cavities \citep{Villenave2019}. Confusion with envelope emission or high inclination could also affect the deficit in the SED. 

Interestingly, the spectral type appears to play a role in the overlap between SED deficit and resolved millimeter cavity. A total of 56 disks have a cavity $\gtrsim$15 au, where the 34 disks that show both a millimeter cavity and a SED deficit are dominated by early type stars with only a quarter of the sample having spectral type K5 or later. The 22 disks that show a millimeter cavity but no SED deficit are dominated by late type stars (K5 - M5), where the only early type stars in this sample have highly inclined disks. Essentially, the probability of missing a transition disk with a cavity $>$15 au based on their SED is 0\% for A-type stars, which increases to 63\% for M-type stars considering the number of M stars with a millimeter cavity without SED deficit (12) compared to the total number of M stars with a millimeter cavity (17). 
As argued before, the current ALMA snapshot surveys likely exclude additional transition disks with large cavities, except for the regions studied at low or non-uniform resolution, Chamaeleon and Taurus. This means that these statistics should be taken with caution as additional cavities may still be discovered in these regions.

For smaller cavities a statistical study is not yet possible, as only a handful of transition disk candidates with (predicted) small inner cavities have been imaged at sufficiently high spatial resolution with ALMA: CX Tau \citep{Facchini2019} and 6 transition disk candidates in Lupus \citep{vanderMarel2022}, but all but one of these disks do not show a small inner cavity but instead, a highly inclined or a very compact dust disk, both of which can explain the small deficit in the SED as well \citep{Birnstiel2012,vanderMarel2022}. Small inner cavities of 1-2 au have been discovered by coincidence in high-resolution ALMA images of XZ Tau B \citep{Osorio2016} and ISO-Oph2B \citep{Gonzalez2020} without deficits in their SED, but as these disks are in multiple systems, the near infrared photometry is likely contaminated by the companions. A handful of $\sim$5-10 au cavities have been resolved as well, see Table \ref{tbl:smallsample}.

The analysis above demonstrates the shortcomings of SED analysis in identifying transition disks: even though SED identified transition disk candidates have a high likelihood to be confirmed with millimeter imaging (for large cavities), a large number of additional transition disks with large cavities have been discovered without SED indicator, which appears to be even more common for late type stars than for early type stars. For smaller cavities ($<$15 au) this issues appears to be even more severe, but only very small sample statistics are available at the moment. This realization is particularly important for studies that use the fraction of SED identified transition disks to derive parameters related to the underlying mechanism, such as time scales of photoevaporative clearing \citep{Ercolano2017} and impact of external vs internal photoevaporation \citep{Coleman2022}. Essentially it is not recommended to use SED-derived statistics for such comparisons, and more complete high-resolution millimeter studies are required to investigate the overall presence of transition disk cavities, large and small. 

\section{The role of transition disks}
\label{sec:analysis}
\subsection{Transition disk demographics}
All resolved transition disks with cavity size $>$15 au within 200 pc are listed in Table \ref{tbl:fullsample} with their main disk and stellar properties using the parallax information from Gaia DR2 \citep{GaiaDR2}. Ages are taken as the average cluster age for T Tauri stars \citep{Michel2021}(1.5$\pm$0.5 Myr for Ophiuchus and Taurus, 2$\pm$0.5 Myr for Chamaeleon, 2.5$\pm$0.5 Myr for Lupus, 5$\pm$0.5 Myr for CrA, 10$\pm$3 Myr for Upper Sco) and individually derived ages for the Herbig stars \citep{Vioque2018}. These ages are  somewhat higher than the average cluster age. All dust masses are calculated using the default assumptions regarding the millimeter flux \citep{Manara2022}. Cavity radii and spectral indices $\alpha_{\rm mm}$ are taken from the literature \citep[e.g.][]{Pinilla2014,Tazzari2020} using a frequency range between 100 and 345 GHz (ALMA Band 3/4 in combination with Band 6/7).  As argued above, this sample can only be considered as complete for cavities $\gtrsim$25 au and possibly incomplete for cavity radii 15-25 au. In addition, Table \ref{tbl:smallsample} lists a handful of resolved transition disks with cavity sizes $<$15 au with their properties, which is very likely incomplete due to lack of high-resolution imaging on the bulk of the disk population, and these disks are not included in the comparison plots. 

Starting from the sample of Table \ref{tbl:fullsample}, the transition disk demographics with respect to  the general protoplanetary disk population are evaluated in this Section, using the data table by \citet{Manara2022} of disk and stellar properties of Class II disks in nearby star forming regions (Taurus, Ophiuchus, Lupus, Chamaeleon, CrA, Upper Sco), which have been corrected for the Gaia DR2 distance as well. 

\begin{figure}[!ht]
\centering
\includegraphics[width=0.9\textwidth]{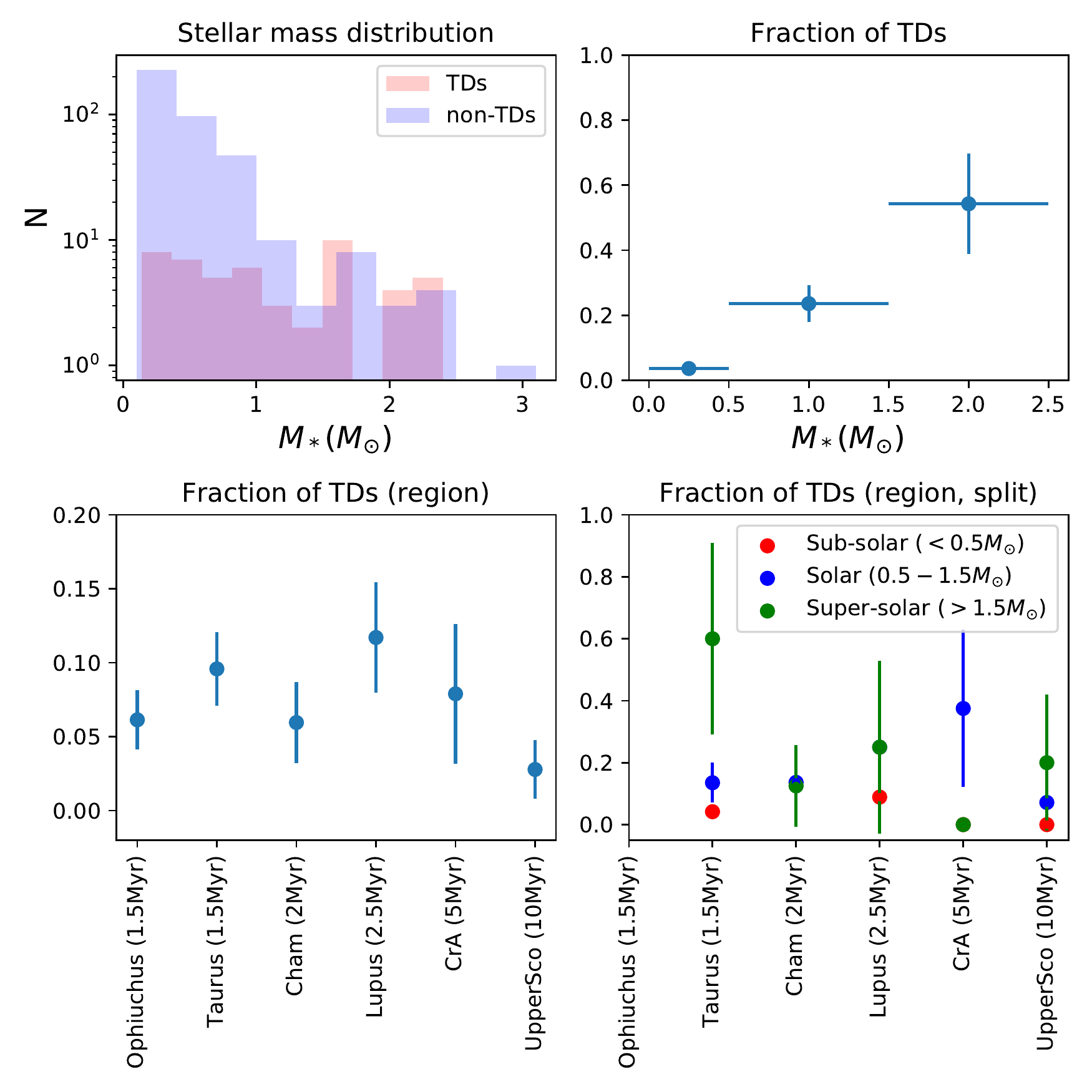}
\caption{Occurrence rates of transition disks. The right panels show the occurrences when split up in sub-Solar ($<$0.5 $M_{\odot}$), Solar (0.5-1.5 $M_{\odot}$) and super-Solar ($>$1.5 $M_{\odot}$), showing that the occurrence increases with stellar mass. For Ophiuchus this assessment cannot be made due to the unknown stellar properties of the bulk of the disk population.}
\label{fig:occurrence}
\end{figure}

Figure \ref{fig:occurrence} displays the general stellar mass distribution and occurrence rates of transition disks per cluster. The stellar mass bins are split up in sub-Solar ($<$0.5 $M_{\odot}$), Solar (0.5-1.5 $M_{\odot}$) and super-Solar ($>$1.5 $M_{\odot}$). The transition disk occurrence increases with stellar mass, as previously derived by \citet{vanderMarelMulders2021}. The overall fraction of transition disks per region is 10+/-3\% for the 4 youngest clusters (fractions are equal within 1$\sigma$), 8$\pm$5\% for the intermediate age cluster CrA and drops to 3$\pm$2\% for Upper Sco. When split up in stellar mass bins, the drop becomes insignificant due to the small number of transition disks in Upper Sco. The overall drop indicates that transition disks, although strong dust traps are thought to be long-lived \citep{Pinilla2020}, eventually dissipate their millimeter dust, which could be either a result of planetesimal growth in the dust rings \citep{Stammler2019,Eriksson2020,Zormpas2022} or general disk dissipation. 

\begin{figure}[!ht]
\centering
\includegraphics[width=\textwidth]{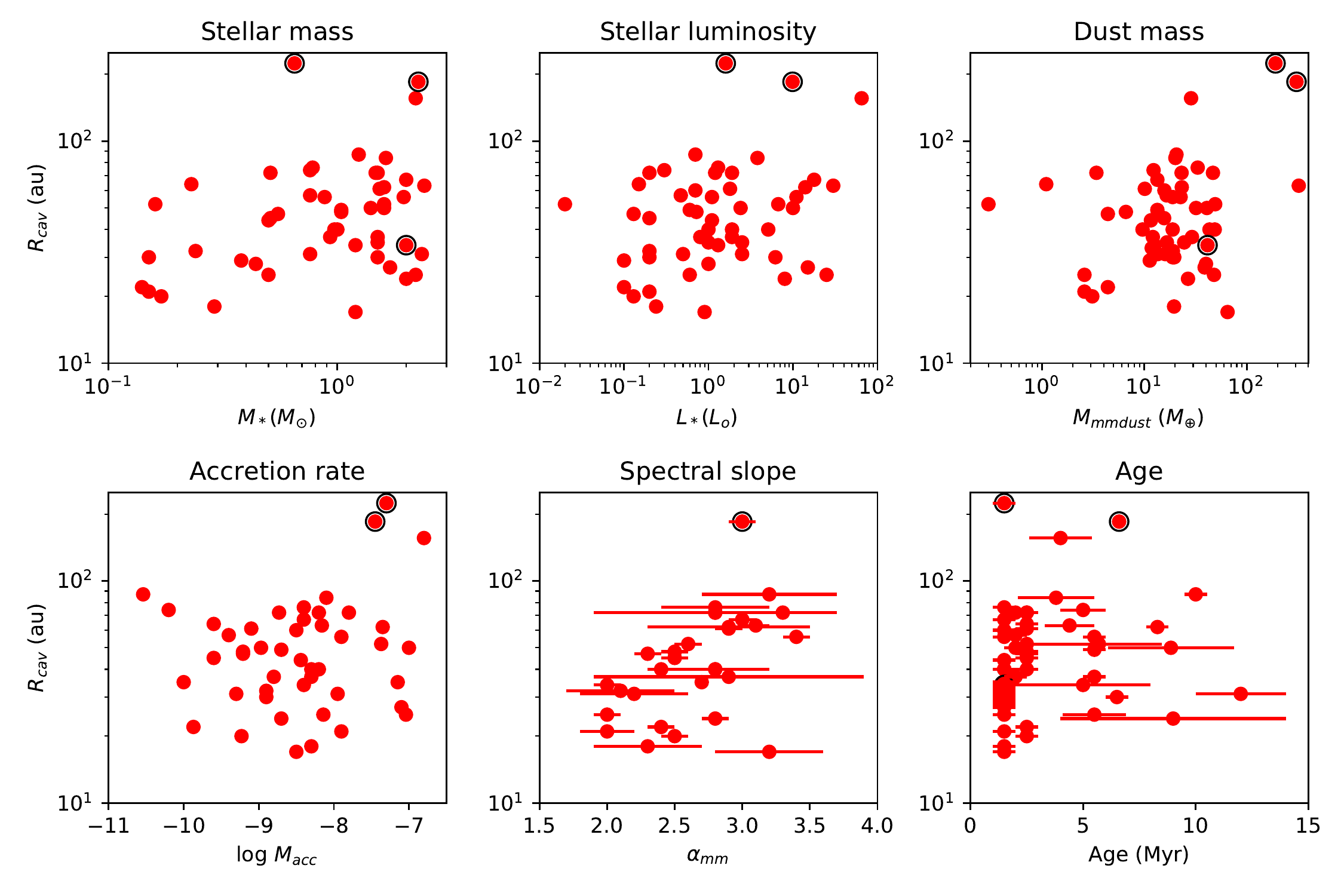}
\caption{Correlation plots between the dust cavity radius and stellar and disk properties: stellar mass, stellar luminosity, dust mass, accretion rate, spectral slope of the millimeter emission and age. The known circumbinary disks with wide orbit companions are encircled and not included in the correlation fit.}
\label{fig:rcav}
\end{figure}

In Figure \ref{fig:rcav} the cavity radius is compared with a number of stellar and disk properties to check for correlations, expanding upon previous studies with smaller samples of transition disks. For each plot, the correlation coefficient is estimated using the \texttt{linmix} package \citep{Kelly2007}, excluding the non-spectroscopic circumbinary disks as their cavity is likely caused by the stellar companion (see Section \ref{sec:companions}). No correlation was found with stellar mass, stellar luminosity, disk dust mass or accretion rate, similar to previous work \citep{Pinilla2018tds}. No trend of cavity sizes is seen with age, although it should be noted that the targets with age $>$5 Myr are dominated by Herbig disks which might distort this picture. A tentative correlation was found with spectral index, consistent with findings by \citet{Pinilla2014}. As the spectral index is higher on average than for typical protoplanetary disks \citep[e.g.][]{Testi2014,Tazzari2020} and it scales with cavity radius, this correlation is interpreted as a lack of optically thick central emission in transition disks \citep{Pinilla2014}. 

The lack of correlation with stellar parameters and disk dust mass implies that the cavity radius (and its underlying clearing mechanism) may not be set by the diversity in initial conditions, i.e. whatever is causing the cavity acts in a similar way and scale across a range of stellar and disk properties, and also remains constant with time considering the lack of trends in the age plot. The implications are further discussed in Section \ref{sec:companions}. The lack of correlation with stellar luminosity argues against any connection with icelines, as previously shown for ring disks \citep{Huang2018,Long2018,vanderMarel2019}. 

\begin{figure}[!ht]
\centering
\includegraphics[width=0.8\textwidth]{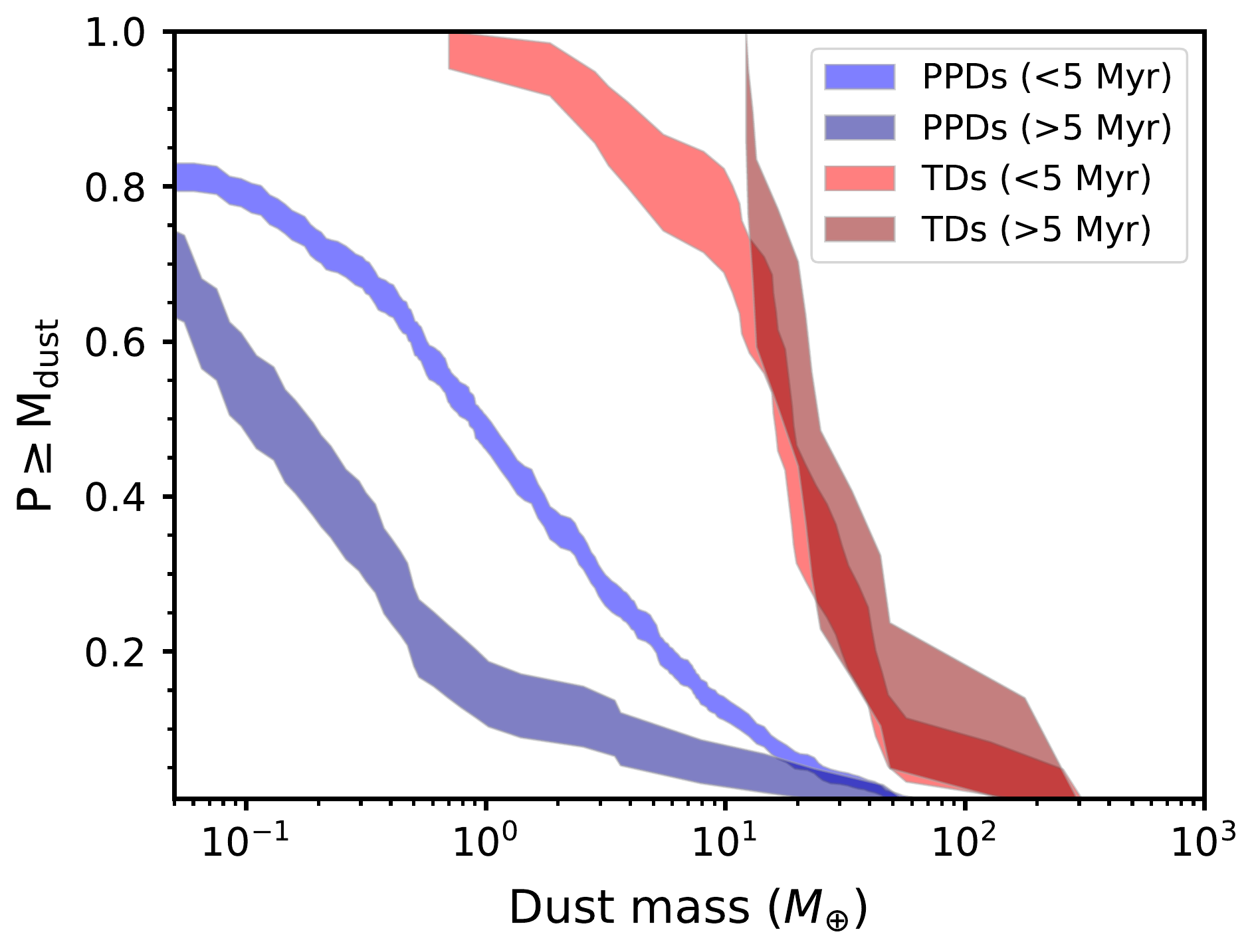}
\caption{Cumulative dust mass distributions of the transition disk sample of this study and the general protoplanetary disk population using the data table from \citet{Manara2022}, indicating the fraction of disks above a given dust mass. The $<$5 Myr group includes the protoplanetary disks from Taurus, Ophiuchus, Lupus, Chamaeleon, whereas the $>$5 Myr group covers the older Upper Sco region and several individual transition disks in e.g. Upper and Lower Cen.}
\label{fig:cmd}
\end{figure}

\begin{figure}[!ht]
\centering
\includegraphics[width=\textwidth]{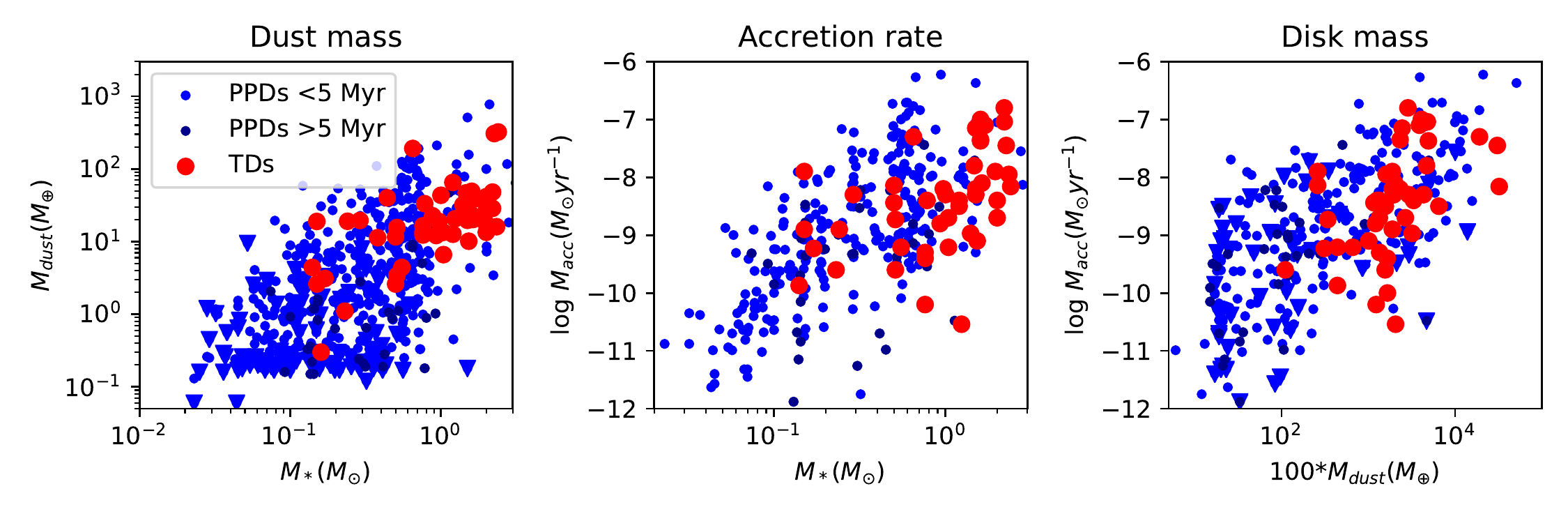}
\caption{Correlation plots between the disk dust mass, stellar mass and accretion rate for both the transition disk sample and the general protoplanetary disk population from \citep{Manara2022}.}
\label{fig:diskpop}
\end{figure}

Figure \ref{fig:cmd} compares the transition disk dust masses with that of the general protoplanetary disk population. It is immediately clear that transition disks have significantly higher dust masses than the general protoplanetary disk population and does not decrease with age, as previously seen \citep{vanderMarel2018,vanderMarelMulders2021}, whereas protoplanetary disk masses do decrease between young and old regions. This plot is demonstrating a trend predicted by \citet{Pinilla2020}, that disks with strong dust traps (such as those in transition disks) maintain their dust mass, whereas disks with no or weak dust traps decrease in dust mass. 

The correlation plots of Figure \ref{fig:diskpop} are less straight-forward to interpret. It has been suggested that transition disks do not follow the general $M_* - M_{\rm dust}$ correlation, but have a much flatter linear slope \citep{Pinilla2018tds}, which would be explained by the presence of strong dust traps in those disks \citep{Pinilla2020}. However, their sample was heavily biased towards higher mass stars, and the more extended transition sample in the current work shows several transition disks around lower mass stars with lower dust masses \citep[e.g.][]{vanderMarel2022}. A linear fit through the logarithmic datapoints gives a slope of 1.0$\pm$0.2 for $M_{\rm dust}$ vs $M_*$ for the transition disks, similar to what is found for the overall disk populations in young clusters \citep{Ansdell2017}. This appears to be consistent with the predictions by \citet{Pinilla2020} for disks with dust traps (see their Figure 7): whereas the real dust mass remains constant and shows a shallow relation with stellar mass, the correlation for the observable millimeter dust mass steepens from 1 to 5 Myr due to more efficient boulder growth in dust traps around low-mass stars \citep[see also e.g.][]{Zormpas2022}. For disks without dust traps, the correlation is already steep at 1 Myr and remains the same in slope and only drops in overall dust mass, due to rapid radial drift. As individual ages remain too uncertain to measure this difference in slope with age, the trends predicted by \citet{Pinilla2020} cannot be directly observed, but the similarity of the two disk populations is generally consistent. 

Studies of accretion rates in transition disks show a mix of conclusions: some studies find the accretion rates to be significantly lower than those of regular protoplanetary disks \citep{Najita2007,Manara2016} whereas other studies find comparable rates \citep{Manara2014}. This discrepancy is  likely caused by the different definitions of transition disks used in these studies. The accretion rates of the transition disk sample with large cavities presented in this work are not significantly different from that of the protoplanetary disk population (Figure \ref{fig:diskpop}): a linear fit through the logarithm of the data points results in $\log(M_{\rm acc} = (-7.9 \pm 0.1) + (1.9 \pm 0.1)\log M_*$ for protoplanetary disks and 
$\log(M_{\rm acc} = (-8.4 \pm 0.2) + (1.6 \pm 0.4) \log M_*$ for transition disks. This implies that the underlying clearing mechanism for these large cavity transition disks may not be related to accretion. In particular, the mechanism that is opening the cavity is not limiting the accretion rate through the cavity, or there is a massive and long-lasting reservoir in their inner disks. This is further discussed in Section \ref{sec:mech}. The third plot in Figure \ref{fig:diskpop} demonstrates that transition disks are on the high end of the disk mass distribution, but not significantly different in terms of accretion rates. 

\subsection{Clearing mechanisms}
\label{sec:clearing}
\subsubsection{Companions}
\label{sec:companions}
The most promising explanation for transition disk cavities is the presence of massive planetary companions that have cleared a gap along their orbit, in particular after the discovery of (accreting) protoplanets of several Jovian masses in PDS~70 and AB~Aur \citep[e.g.][]{Keppler2018,Haffert2019,Currie2022}. The accretion rate can remain high in the presence of a giant planet that has cleared a gap \citep{LubowdAngelo2006}, and this accretion has been suggested to be wind-driven \citep{WangGoodman2017,Rosenthal2020,Martel2022}. Furthermore, the stellar mass dependence of transition disks corresponds to that of Jovian exoplanets at smaller orbital radii of 3-8 au, suggesting that the planets which potentially open the cavities would eventually migrate inwards \citep{vanderMarelMulders2021}.

Before these discoveries, the idea of massive Jovian planets in cavities was met with skepticism due to the large orbital radii of several tens of au that were required, inconsistent with both planet formation models \citep[e.g.][]{Pollack1996,Helled2014,Drazkowska2022}, difficulties of opening a wide gap with a single planet \citep{Zhu2011}, and with direct imaging studies around main sequence stars indicating that such companions were rare \citep[see][for an occurrence analysis]{Nielsen2019,Vigan2021}. This has led to the possibility that multiple companions must be responsible for the cavities \citep{Dodson-Robinson2011}. High contrast imaging studies resulted in several debates on claims of companion candidates in transition disk cavities \citep[e.g.][]{Huelamo2011,Biller2014,Sallum2015,Currie2015,Thalmann2016,Rameau2017,Ligi2018,Reggiani2018,Currie2019,Wagner2019}. Overviews of detection limits have been published for several transition disk samples, although the detection limit varies strongly with radius and from disk to disk, the conversion to planet mass is model-dependent and dependent on the assumed circumplanetary disk  \citep{Zhu2015,Brittain2020,vanderMarel2021asymm,Asensio2021}. Also H$\alpha$ imaging has mostly resulted in upper limits \citep{Zurlo2020,Close2020}. Overall it has become clear that current high contrast imaging instruments are limited in their detection possibilities down to several Jovian masses, especially in the inner part of the cavity \citep{Benisty2022,Currie2022}. It has also been suggested that the detection of the protoplanets in PDS70 may be the result of the protoplanets being in an early stage of accretion, whereas protoplanets in other transition disks have already finished their accretion and are naturally fainter \citep{Francis2020,Brittain2020}. As the possibilities and limitations of high contrast imaging are discussed in detail in these review chapters, and as limits on companions are best treated on an individual basis, this is not further discussed here. 

Although circumbinary disks have historically been proposed to explain some transition disks, such as for Coku/Tau 4 \citep{Ireland2008}, for many other transition disks in Taurus and Ophiuchus a brown dwarf companion  ($>$20 $M_{\rm Jup}$) in the inner $\sim$40 au has been excluded by sparse aperture masking \citep{Kraus2008,Kraus2011,Cheetham2015,Ruiz2016} and Keck interferometry \citep{Pott2010}, as well as spectroscopic binary searches \citep{Kohn2016}.. It should be noted that Coku/Tau 4, despite being a famous circumbinary transition disk, has actually not been spatially resolved yet with millimeter interferometry. Other transition disks have been proposed to be circumbinary based on modeling indirect signatures in the cavity such as spirals and warps \citep{Calcino2019,Poblete2020,Poblete2022,Norfolk2022}. Spectroscopic binaries have been discovered as well in some transition disk cavities, but their separations are usually well within 1 au and thus insufficient to explain a cavity $>$20 au in radius. Table \ref{tbl:circumbinary} lists all transition disks within 200 pc with known inner stellar companions and limits on those. 

The key contribution of ALMA in the question on the origin of cavities has been the spatial mapping of the gas distribution in the cavities, as explained in Section \ref{sec:gasdust}. The deep gas cavities that are often at least a factor 1.5 times smaller than the dust cavities point towards super-Jovian planets at orbital radii $\sim$10-20 au inside dust cavities of 40-80 au \citep[e.g.][]{Pinilla2012b,vanderMarel2016-isot,Facchini2018gaps,vanderMarel2021asymm}. The resulting large dust cavity radii can be explained by the presence of multiple planets such as seen in PDS~70, with a derived gas cavity radius of 22 au \citep{vanderMarel2021asymm}, a dust ring radius of 74 au with an inner shoulder down to 45 au and protoplanets at 20 and 35 au \citep{Keppler2018,Haffert2019}. The dust ring radius can be considered as the peak of the pressure bump at the outer edge of the gas gap. In AB~Aur, the protoplanet was found at 93 au \citep{Currie2022}, with a gas cavity radius of 98 au \citep{vanderMarel2021asymm} and a dust ring radius at 170 au again with an inner shoulder. For other transition disks, an inner dust shoulder is not always detected, possibly due to lower sensitivity and resolution, and the dust ring would be equivalent to the dust cavity radius. A large separation between protoplanet and dust ring radius has also been suggested to be caused by eccentric cavities which develop naturally for super-Jovian planets \citep{Kley2006,Muley2019}. An eccentric cavity has been measured for the transition disk in MWC~758 \citep{Dong2018obs} but eccentricity is generally unconstrained.

\subsubsection{Other clearing mechanisms}
\label{sec:mech}

\begin{figure}[!ht]
\includegraphics[width=0.9\textwidth]{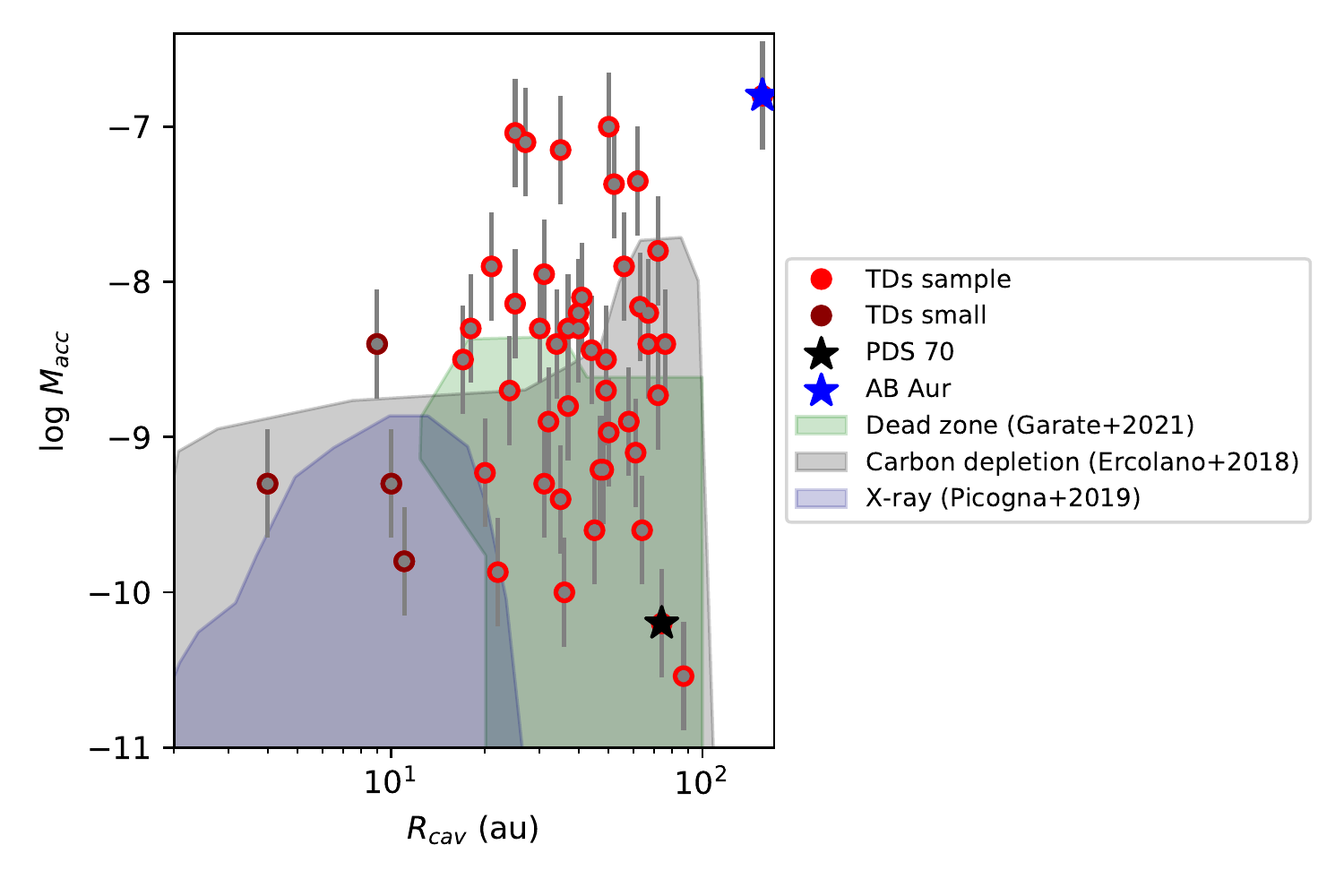}
\caption{Relation between cavity radius and accretion rate which can be reproduced by various photoevaporation models \citep{Ercolano2018,Picogna2019,Garate2021} with the data points of our transition disk sample plus the resolved transition disks with small cavities. The cavities caused by the protoplanets of PDS~70 and AB~Aur are overplotted as well. Figure based on \citet{vanderMarel2022}.}
\label{fig:photoevap}
\end{figure}

A second mechanism to explain the presence of transition disk cavities is photoevaporative clearing at the end of the lifetime of the disk \citep{Alexander2014}, when the accretion rate drops below the photoevaporative rate and a deep gap is rapidly created, clearing the disk from the inside out \citep{Clarke2001}. This explanation would connect naturally with the term `transition disk' as an evolutionary state and transition disk fractions have been used to determine the expected timescale of the clearing \citep[e.g.][]{Ercolano2017}. However, the non-uniform definitions and uncertainties in SED-derived transition disk classifications challenges this approach \citep{vanderMarel2022}. Whereas in initial models photoevaporation could only explain the transition disks with small cavities $<$30 au and low accretion rates $<10^{-9} M_{\odot}$ yr$^{-1}$ \citep{OwenClarke2012,Ercolano2017}, recent efforts in developing more advanced prescriptions of photoevaporation, e.g. including X-rays, carbon depletion and dead zones in the inner disk \citep{Ercolano2018,Wolfer2019,Picogna2019,Garate2021} can reproduce a much larger range of transition disk cavities, although not all of them (see Figure \ref{fig:photoevap}). In contrast to previous comparisons, no SED derived cavities are included in this plot as their transition disk nature remains debatable \citep{vanderMarel2022}. On the other hand, the cavity of PDS~70 can be reproduced as well by this model while it is known to contain two protoplanets. Whether photoevaporation can indeed be a feasible explanation for observed transition disks thus remains a matter of debate until more firm constraints on protoplanet detections have been obtained. Also the direct detection of photoevaporative winds would constrain such a mechanism (for a full discussion on disk winds see \citet{Pascucci2022}).

Other mechanisms for transition disk cavities have been proposed as well, although with limited observable predictions. So-called dead zones (regions of low ionization where the magnetorotational instability is inhibited) are capable of creating a millimeter dust ring or asymmetry \citep{Regaly2012,Flock2015} but are not consistent with the deep gas cavities in transition disks \citep{vanderMarel2016-isot}, unless combined with inner clearing by a MHD disk wind \citep{Pinilla2016-dz}. Detailed modeling indicates that pressure bumps at dead zone edge may not necessarily develop \citep{Delage2022}. Enhanced dust growth was shown to lead to a deficit in the SED \citep{Dullemond2005}, but a complete removal of the millimeter grains in the inner part of the disk leading to a spatially resolved millimeter-dust cavity is unlikely considering continuous fragmentation \citep{Birnstiel2012}. However, this mechanism could perhaps explain the very compact millimeter dusk disks \citep{Facchini2019,vanderMarel2022}. More generally, a range of hydrodynamic instabilities have been proposed to explain the presence of gaps and rings such as those seen in the DSHARP survey, which might be able to explain some of the transition disk cavities as well (see discussion in \citet{Andrews2020}).

\subsection{Disk chemistry}
\label{sec:diskchem}
The disk chemical composition sets the atmospheric composition of planets forming in these disks \citep{ObergBergin2021}. Transition disks are particularly interesting in this context, due to their possible prevalence of giant planets in these disks, and also because of their irradiated cavity wall, which may lead to increased evaporation of complex organic molecules that are otherwise locked up in ices \citep{Cleeves2011}. Observationally, transition disks are often included in samples of disk chemistry studies without showing significant differences with other disks \citep[e.g.][]{Oberg2011,Oberg2021}. However, two transition disks stand out in this regard: both HD~100546 and Oph~IRS48 show a rich complex organic chemistry \citep{Booth2021a,Booth2021b,vanderMarel2014,vanderMarel2021-irs48,Brunken2022} with several molecular detections, in particular methanol, CH$_3$OH. The latter was surprising as both disks are Herbigs which are too warm to contain a large amount of CO-ice, which is required to form significant amounts of CH$_3$OH in disks \citep{Herbst2009}. This has been interpreted as inheritance of ices from the dark cloud phase into the protoplanetary disk phase where the ices can be released as pebbles drift inwards \citep{Booth2021a}. CH$_3$OH has also been detected in TW Hya \citep{Walsh2016} but in much smaller abundance and thus unlikely to be related to ice traps. 

\begin{figure}[!ht]
\centering
\includegraphics[width=0.6\textwidth]{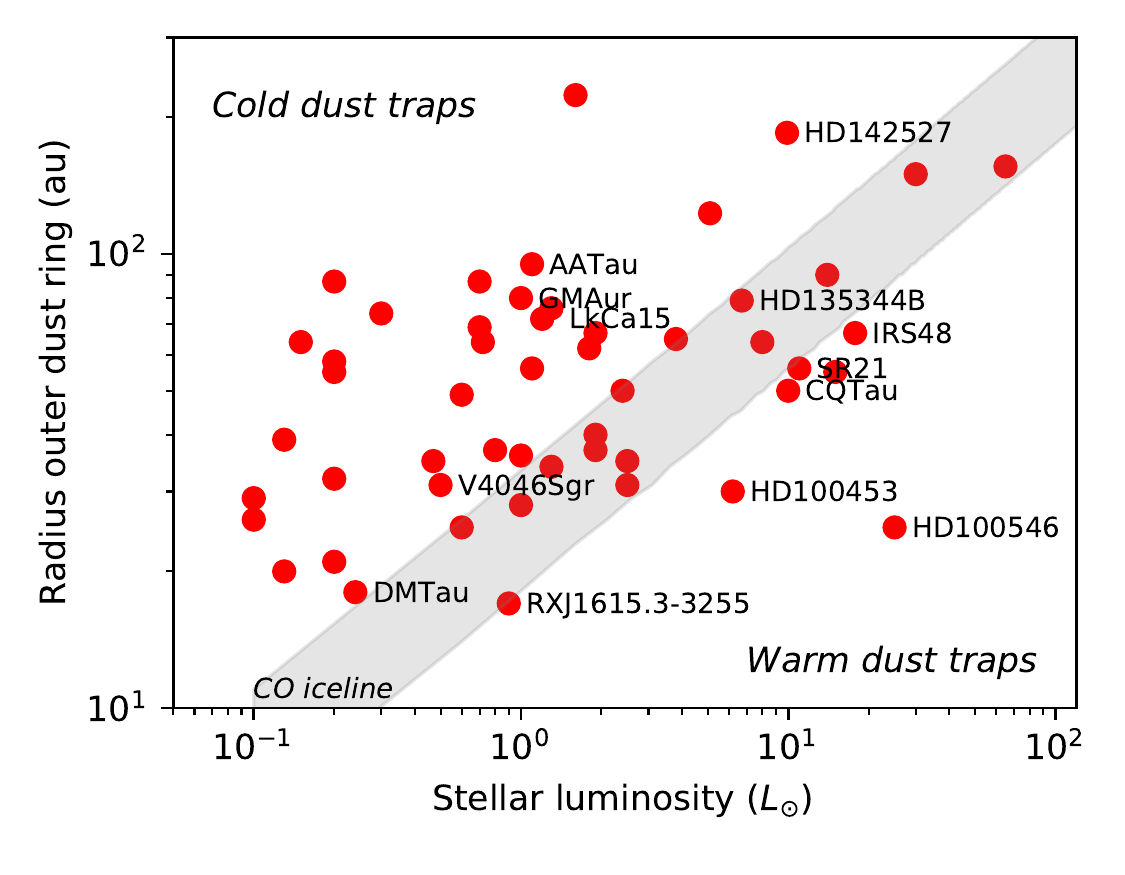}
\caption{Distribution of the outer dust rings in transition disks and the stellar luminosity. The grey shaded area indicates the 22-30 K regime in the midplane as expected from the stellar luminosity, which corresponds to the typical CO freeze out temperature. Disks that are warm enough to evaporate their ices and disks that have been included in disk chemistry surveys \citep{Oberg2011,Oberg2021} are marked by name.}
\label{fig:icetrap}
\end{figure}

The question is why only these two transition disks show such a rich chemistry compared to other disks. The answer may lie in the unique combination of the cavity edge location and the temperature gradient in the disk: the cavity edge happens to lie in the warmer $>$30 K regime of the disk, and gas temperatures can reach $>$100 K in the layers above the midplane \citep{vanderMarel2021-irs48}. As dust pebbles drift inwards towards the nearest dust trap, they transport their icy layers as well \citep{Booth2019,Krijt2020}. If that dust trap (cavity edge) happens to be warm enough, the icy layers can be released. As the disk mid-plane temperature is set by the stellar luminosity following $T(r) \propto \sqrt[4]{L_*/r}$, the expected 30 K line can be compared with the dust trap locations in transition disks. However, as some transition disks have outer dust rings as well, it is important to consider the location of the most \emph{outer} dust trap, not necessarily the cavity edge. Figure \ref{fig:icetrap} shows this comparison for the transition disks considered in this study. The majority of the transition disks (`cold dust traps') lie above the expected CO iceline (22-30 K) \citep{Bisschop2006, Noble2012}, i.e. their dust traps are likely too cold to have their icy content evaporated, even above the midplane. This sample includes several disks that have been studied for chemical complexity. On the other hand, the `warm dust traps' (including IRS48 and HD100546 where CH$_3$OH has been detected) are likely warm enough for evaporation and are thus key targets to search for chemical complexity.

The link between dust transport and disk composition has been suggested independently in e.g. the anti-correlation of warm H$_2$O emission with disk dust size \citep{Banzatti2020,Kalyaan2021}, rovibrational emission in TW~Hya \citep{Bosman2019}, chemical composition of accreting material onto the star  \citep{McClure2020} and in the bimodal distribution in C$_2$H emission, a tracer of the elemental C/O ratio \citep{vanderMarel2021-c2h}. Transition disks are part of these samples and the comparison of disks with and without strong dust traps is key for our understanding of the influence of dust transport on disk composition. 

\subsection{Fate of transition disks}
\label{sec:debris}
It has been suggested that disks with dust traps, including transition disks, are the progenitors of the debris disks \citep{Cieza2020,Michel2021,vanderMarelMulders2021}, i.e. the planetesimal belts around main sequence stars $>$10 Myr old, based on a dust evolution argument: only if sufficient pebbles are trapped at large orbital radii they can grow to sufficiently massive planetesimal belts that produce the debris observed at infrared and submillimeter wavelengths \citep[e.g.][]{Hughes2018}. When unhindered by pressure bumps, pebbles would drift inwards before having the time to grow planetesimals \citep{Pinilla2020}. Modeling work shows that this is indeed quantitatively feasible considering debris and protoplanetary disk infrared luminosities and time scales \citep{Najita2022}.

\begin{figure}[!ht]
\centering
\includegraphics[width=0.8\textwidth]{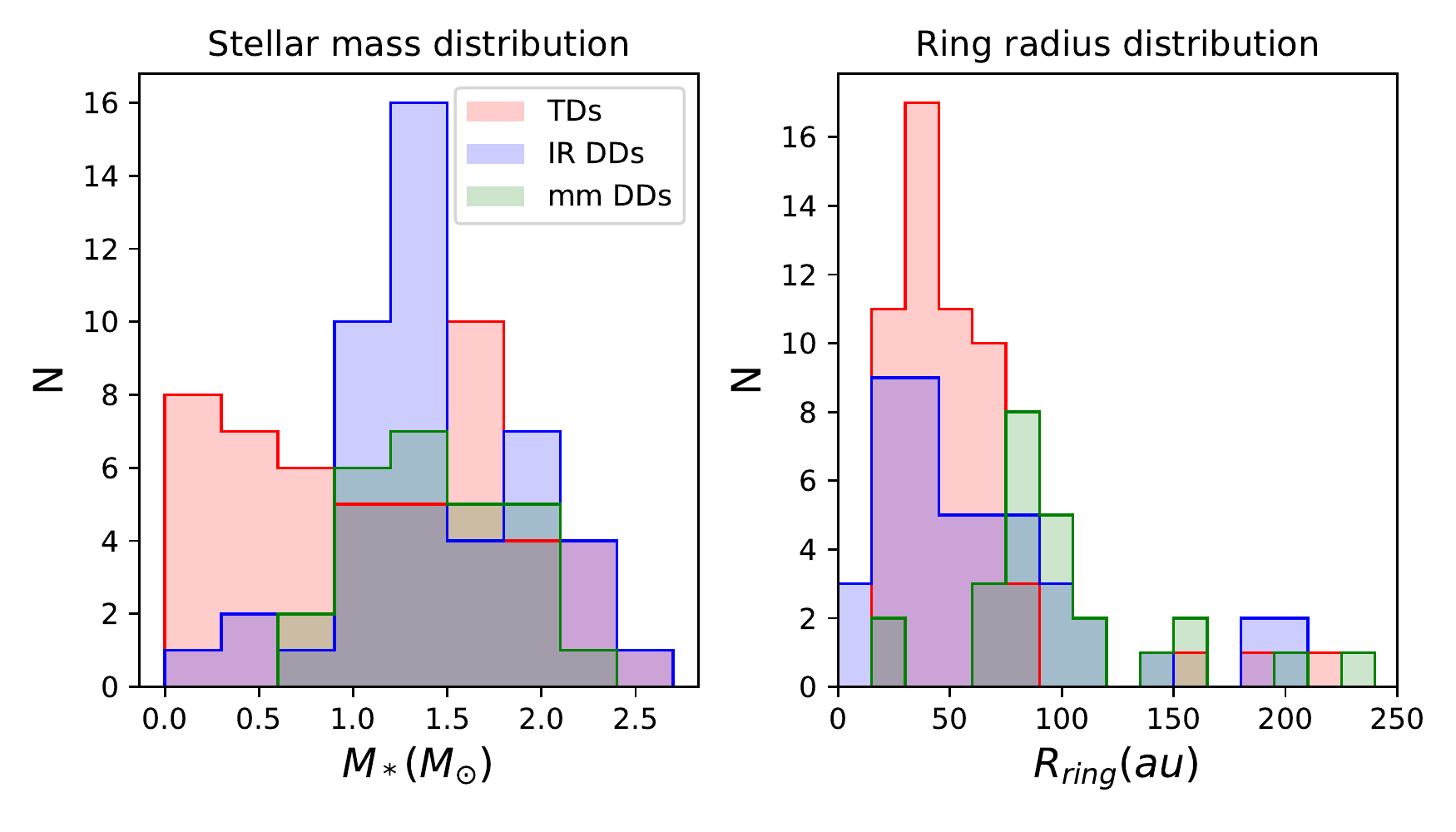}
\caption{Distribution of stellar masses and dust ring locations in the transition disk sample and in the debris disk samples from the \emph{Herschel} SONS survey \citep{Holland2017} and the millimeter ALMA sample \citep{Matra2018}.}
\label{fig:debrisdisks}
\end{figure}

Figure \ref{fig:debrisdisks} shows the distribution of stellar masses and ring radii of transition disks vs debris disks, using both the cold debris disk sample from the \emph{Herschel} SONS survey \citep{Holland2017} and the millimeter debris disk sample from the ALMA sample \citep{Matra2018}. Although transition disks appear to be more often detected around lower mass stars than debris disks, this is likely a selection effect, as the absolute number of low-mass stars is much higher in the IMF. Both the debris disk and transition disk populations show an increasing occurrence with stellar mass supporting the hypothesis that they are connected \citep{Sibthorpe2018,vanderMarelMulders2021}. It is possible that even more transition disks around low-mass stars are to be discovered as these disks are generally fainter and not yet imaged at high angular resolution. The ring radii of both populations are comparable for the SONS survey, whereas the ALMA debris disk survey distributions peaks at somewhat larger dust ring radii, but this can be a selection bias as this sample was not uniformly selected \citep{Matra2018}. Overall, this demonstrates that the dust ring radii are indeed comparable. Also the discovery of so-called hybrid disks is interesting in the evolutionary context: these disks, such as HD141569, have a significant amount of gas compared to their dust content, and are thought to be in transition between protoplanetary and debris disks \citep[e.g.][]{Pericaud2017,Miley2018}. The dust and gas morphology of HD141569 has been suggested to be ring-like as well \citep{Currie2016,White2018,diFolco2020}, consistent with transition disks.

\begin{figure}[!ht]
\centering
\includegraphics[width=0.6\textwidth]{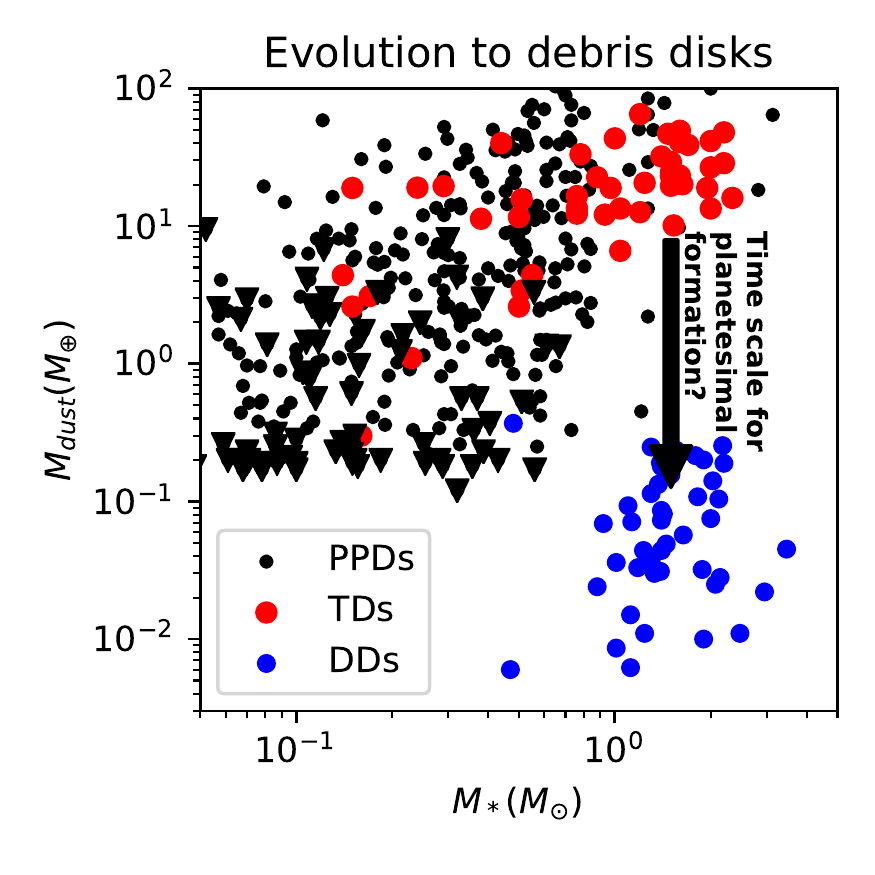}
\caption{The $M_* - M_{\rm dust}$ correlation of protoplanetary disks, transition disks and debris disks based on the data from Taurus, Ophiuchus, Lupus and Chamaeleon from \citet{Manara2022} and the debris disk data from \citet{Holland2017}.}
\label{fig:debrisdiskevol}
\end{figure}

Debris disk millimeter dust masses are several orders of magnitude below that of protoplanetary disks \citep{Lovell2021,Michel2021}. The millimeter dust in debris disks is secondary, i.e. it is generated by a collisions of planetesimals \citep{Wyatt2008}. Figure \ref{fig:debrisdiskevol} compares the $M_* - M_{\rm dust}$ correlation of protoplanetary disks, transition disks and debris disks. If the transition disks are the progenitors of debris disks, their observable dust mass has to decrease by 2-3 orders of magnitude for super-Solar stars. For sub-Solar masses the statistics of debris disk masses are insufficient. Dust evolution models of protoplanetary disks including boulder formation already predict a decrease of 1 orders of magnitude of obserable dust mass in disks with dust traps \citep{Pinilla2020,Zormpas2022}. It remains a question whether such faint protoplanetary disk rings exist, and what that tells about the timescales for boulder and planetesimal formation. The evolved Class III disks with moderate infrared excess are good candidates, but they have not been studied uniformly so far \citep{Michel2021}. One difficulty of studying Class III disks is the cluster membership, which has only recently been resolved with Gaia DR2: for example, the Lupus Class III survey by \citep{Lovell2021} is likely contaminated with a large number of objects from the much older UCL region \citep{Michel2021}. Also the inferred dust size distribution and resulting opacity of protoplanetary and debris disks remains highly uncertain \citep{Manara2022}. More dedicated surveys of evolved disks are required to bridge the gap between protoplanetary and debris disks. 

\section{Summary and outlook}
\label{sec:discussion}
The spatially resolved imaging of transition disks with large cavities has revolutionized our understanding of the physical mechanisms behind them. The direct and indirect evidence for protoplanets embedded in their cavities is piling up, providing answers to one of the major questions of the previous decade regarding transition disks: can massive planets at large orbits be responsible for the cavities? At the same time photoevaporation and dead zone models have continued to develop and the question whether some transition disks may ben caused by something other than companions remains open. In context of companions, the interest within the field has somewhat shifted from transition disks to disks with gaps in general, but it is clear that transition disks still provide great test beds for models of planet disk interaction. 

Further characterizations and detections (or deeper limits) of giant planets in transition disk cavities are crucial for determining the general nature of the cavity clearing. Direct imaging with new facilities such as \emph{James Webb Space Telescopes}, the upgraded the VLT/SPHERE, Gemini/GPI and Subaru/SCExAO instruments and in the future the 30m-class telescopes will provide much deeper limits on embedded protoplanets, in particular at closer distances to the host star. Also opportunities of circumplanetary disk detection at both millimeter and infrared wavelengths are promising.

Clearly, the confirmation of planetary companions at wide orbits creates an issue for giant planet formation models, as it requires an efficient (and rapid) formation process. Considering the existence of gapped disks even at 1-2 Myr, this implies that giant planet cores with sufficient mass to create a pressure bump form within that timescale, and have already accreted enough gaseous material to clear a wide, deep gap in the gas. At the same time, inward migration must be halted or slowed down sufficiently to explain how large scale gaps and cavities exist for several Myr, and the accretion onto the host star has to be maintained in the presence of deep gas gaps. Planet formation and planet-disk interaction models will have to be able to reproduce this to explain the observations. One important insight here is the finding that transition disks tend to be massive and more often found around higher mass stars, implying that high initial disk masses have to be used as input for such formation models. However, this could be the result of an observational bias towards brighter disks and cannot be confirmed until deep high-resolution ALMA observations of the majority of the smaller, fainter protoplanetary disks are obtained.

On the other hand, the efficiency of dust evolution, growth and transport in pressure bumps remains an active topic of study. Are the lower dust masses seen in transition disks around low-mass stars the result of dust growth to planetesimals, or lower initial disk masses? Can transition disk rings indeed be the progenitors of debris disks, and if so, how can we prove this? Can dust transport indeed change the chemical composition across the disk? How does the dust in the inner disk evolve and how is it related to the flow of material through the gap? Is dust growth different in azimuthal dust traps than in radial dust traps? Can the azimuthal nature of some dust traps indeed be explained by their Stokes number? Considering optical depth issues at millimeter wavelengths, spatially resolved centimeter observations tracing optically thin larger grains will provide more insights, which will be possible with the low frequency receivers of ALMA Band 1 and 2, and with the future ngVLA. Chemical composition of the inner disk as measured with e.g. \emph{JWST} in comparison with the outer disk as measured with ALMA will provide more answers on the transport of ices and the consequences for exoplanet atmosphere composition.

One of the major open questions in transition disk studies is whether smaller cavities $<$10 au indeed exist as well in millimeter wavelengths, and if so, whether these are simply scaled down versions of the spatially resolved large cavities. The lack of high-resolution ALMA observations of smaller disks is a clear limitation to our interpretation of the disk population and the role of transition disks, in particular with respect to the small cavity transition disk population identified by their SEDs. ALMA certainly has the capacity to perform such imaging campaigns, and it is time to shift the focus from brighter, larger disks to the smaller disks to fully reveal the smaller scale substructures and their underlying mechanisms. 

\backmatter





\bmhead{Acknowledgments}
I would like to thank the anonymous referee for their constructive report and useful suggestions. Furthermore, I would like to thank Jonathan Williams, Sean Andrews, Richard Alexander, Antonio Garufi and Stefan Kraus for their suggestions to improve the manuscript. 


\newpage

\begin{appendices}

\section{Data tables}\label{tables}
This Appendix contains all tables used in the analysis of this work.

\begin{table}[!ht]
\caption{SED transition disk candidates}
\label{tbl:sedcandidates}
\resizebox{\textwidth}{!}{%
\begin{tabular}{lllllll}
\hline
\multicolumn{7}{c}{\emph{SED identified transition disks and their millimeter cavity status}}\\
\hline
Region & Target & $R_{\rm cav}$(au) & Status$^a$ & SpT & Ref $^b$ & Notes \\
\hline
Chamaeleon	&	CSCha	&	37	&	Y	&	K2	&1&		\\	
Chamaeleon	&	HD97048	&	63	&	Y	&	A0	&1&		\\	
Chamaeleon	&	SZCha	&	72	&	Y	&	K0	& 2&	cavity size corrected for Gaia DR2 	\\	
Chamaeleon	&	CHXR22E	&	-	&	U	&	M3.5	&3&	no cavity at 0.6” resolution	\\ 
Chamaeleon	&	ESOH$\alpha$559	&	-	&	U	&	M5	&4&	no cavity at 0.6” resolution	\\ 
Chamaeleon	&	ISO-ChaI-52	&	-	&	U	&	M4.5	&4&	no cavity at 0.6” resolution	\\ 
Chamaeleon	&	T25	&	-	&	U	&	M4	& - &		\\ 
Chamaeleon	&	T35	&	-	&	U	&	K5	&4&	no cavity at 0.6” resolution	\\ 
CrA	&	RXJ1842.9-3532	&	37	&	Y	&	K2	&1&		\\	
CrA	&	RXJ1852.3-3700	&	49	&	Y	&	K2	&1&		\\	
CrA	&	CrA-9	&	-	&	U	&	M1	&5&	no cavity at 0.3” resolution	\\ 
EpsiCha	&	Tcha	&	34	&	Y	&	G8	&1&		\\	
Isolated	&	HD169142	&	26	&	Y	&	A5	&1 & \\			
LowerCen	&	HD100453	&	30	&	Y	&	F0	&1 & \\			
LowerCen	&	HD100546	&	27	&	Y	&	A0	&1 & \\			
Lupus	&	J16083070-3828268	&	62	&	Y	&	K2	& 6,7 &		\\	
Lupus	&	J16102955-3922144	&	11	&	Y	&	M4.5	&6,7&		\\	
Lupus	&	Sz111	&	55	&	Y	&	M1	&6,7&		\\	
Lupus	&	Sz84	&	12	&	Y	&	M5	&6,7&		\\	
Lupus	&	Sz91	&	86	&	Y	&	M1	&1&		\\	
Lupus	&	MYLup	&	-	&	N	&	K0	&8&	no cavity at 0.04”, high inclination	\\	
Ophiuchus	&	DoAr44	&	40	&	Y	&	K2	&1&		\\	
Ophiuchus	&	IRAS16201-2410	&	28	&	Y	&	G0	&9&		\\ 
Ophiuchus	&	IRS48	&	83	&	Y	&	A0	&1&		\\	
Ophiuchus	&	RXJ1615.3-3255	&	17	&	Y	&	K5	&10&	cavity size corrected for Gaia DR2	\\	
Ophiuchus	&	RXJ1633.9-2442	&	36	&	Y	&	K7	&11&		\\	
Ophiuchus	&	SR21	&	56	&	Y	&	G4	&1&		\\	
Ophiuchus	&	WSB60	&	32	&	Y	&	M6	&1&		\\	
Ophiuchus	&	DoAr28	&	-	&	U	&	K5	&-&		\\	
Ophiuchus	&	J16281385-2456113	&	-	&	U	&	K7	&-&		\\	
Ophiuchus	&	J162740.3-242204	&	-	&	N	&	K5	&12&	no cavity at 0.25” resolution	\\	
Ophiuchus	&	J163023.4-245416	&	-	&	N	&	-	&12&	no cavity at 0.25” resolution	\\	
Ophiuchus	&	J163115.7-243402	&	-	&	N	&	-	&12&	no cavity at 0.25” resolution	\\	
Taurus	&	ABAur	&	156	&	Y	&	A0	&1&		\\	
Taurus	&	CQTau	&	50	&	Y	&	F2	&1&		\\	
Taurus	&	DMTau	&	25	&	Y	&	M2	&1&		\\	
Taurus	&	GMAur	&	40	&	Y	&	K5	&1&		\\	
Taurus	&	IPTau	&	25	&	Y	&	M0	&1&		\\	
Taurus	&	LkCa15	&	76	&	Y	&	K2	&1&		\\	
Taurus	&	MWC758	&	62	&	Y	&	A7	&1&		\\	
Taurus	&	UXTauA	&	31	&	Y	&	G8	&1&		\\	
Taurus	&	V892Tau	&	27	&	Y	&	A0	&13 & \\			
Taurus	&	DETau	&	-	&	U	&	M3	& -&		\\	
Taurus	&	IRAS04125+2902	&	-	&	U	&	M0	&-&		\\	
Taurus	&	J04184133+2827250	&	-	&	U	&	K7	&-&		\\	
Taurus	&	J04324911+2253027	&	-	&	U	&	K5.5	&-&		\\	
Taurus	&	CIDA7	&	-	&	N	&	M5	&14&	no cavity at 0.1” resolution	\\	
Taurus	&	V410X-ray6	&	-	&	N	&	M4.5	&15&	non-detection	\\	
Upper Sco	&	HD143006	&	41	&	Y	&	G5	&8&		\\	
Upper Sco	&	J160421.7-213028	&	87	&	Y	&	K3	&1&		\\	
UpperCen	&	HD135344B	&	52	&	Y	&	F5	&1&		\\	
UpperCen	&	HD142527	&	185	&	Y	&	F6	&1&		\\	
UpperCen	&	HD139614	&	-	&	U	&	A9	&16&	no cavity at 0.5” resolution	\\

\hline
\end{tabular}}
$^a$ Y = cavity confirmed by millimeter imaging; N = no cavity found at millimeter imaging (see last column for resolution); U = unknown, millimeter imaging insufficient to confirm or non-existent. \\
$^b$ References millimeter cavity size. 1) \citet{Francis2020}; 2) \citet{Pinilla2018tds}; 3) \citet{Long2018}; 4) \citet{Pascucci2016}; 5) \citet{Cazzoletti2019}; 6) \citet{vanderMarel2018}; 7) \citet{vanderMarel2022}; 8) \citet{Andrews2018}; 9) \citet{Cieza2020}; 10) \citet{vanderMarel2015-12co}; 11) \citet{Cieza2020}; 12) \citet{Cieza2019}; 13) \citet{Long2021}; 14) \citet{Kurtovic2021}; 15) \citet{WardDuong2018}; 16) \citet{Stapper2022}.
\end{table}

\begin{table}[!ht]
\caption{New millimeter resolved transition disks}
\label{tbl:newtds}
\begin{tabular}{llllll}
\hline
\hline
\multicolumn{6}{c}{\emph{Disks with millimeter cavity without previous SED identification}}\\
\hline
Region	&	Target	&	$R_{\rm cav}$(au)	&&	SpT	&	Ref $^a$	\\
\hline
Chamaeleon	&	HPCha	&	50	&&	K7	& 1\\
Chamaeleon	&	SYCha	&	35	&&	K5	&2\\
CrA	&	PDS99	&	56	&&	K6	&1\\
Isolated	&	V4046Sgr	&	31	&&	K7	&1\\
Lupus	&	J16070384	&	52	&&	M4.5	&3\\
Lupus	&	J16070854-3914075	&	40	&&	M5	&3\\
Lupus	&	J16090141	&	64	&&	M4	&3\\
Lupus	&	RYLup	&	67	&&	K2	&3\\
Lupus	&	Sz100	&	26	&&	M5.5	&3\\
Lupus	&	Sz118	&	64	&&	K5	&3\\
Lupus	&	Sz123A	&	39	&&	M1	&3\\
Lupus	&	Sz129	&	10	&&	K7	&4\\
Ophiuchus	&	ROXRA3/rho-Oph2	&	69	&&	M0	&5\\
Ophiuchus	&	SR24S	&	35	&&	K6	&1\\
Ophiuchus	&	WSB82/rho-Oph38	&	50	&&	K0	&5\\
Taurus	&	AATau	&	44	&&	K5	&1\\
Taurus	&	CIDA1	&	21	&&	M4.5	&6\\
Taurus	&	CIDA9A	&	29	&&	M2	&1\\
Taurus	&	GGTau	&	224	&&	K7	&1\\
Taurus	&	MHO2	&	28	&&	M3	&1\\
Taurus	&	MHO6	&	10	&&	M5	&6\\
Taurus	&	RYTau	&	27	&&	G2	&1\\
Taurus	&	UZTau	&	10	&&	M2	&7\\
Taurus	&	ZZTauIRS	&	50	&&	M5	&8\\
UpperCen	&	PDS70	&	74	&&	K7	&1\\

\hline
\hline
\end{tabular}\\
$^a$ References millimeter cavity size. 1) \citet{Francis2020}; 2) \citet{Orihara2022}; 3) \citet{vanderMarel2022}; 4) \citet{Andrews2018}; 5) \citet{Cieza2020}; 6) \citet{Kurtovic2021}; 7) \citet{Long2018}; 8) \citet{Hashimoto2021b}.
\end{table}

\begin{table}[!ht]
\caption{Full sample of transition disks within 200 pc with cavities $>$15 au}
\label{tbl:fullsample}
\resizebox{\textwidth}{!}{%
\begin{tabular}{lllllllllllll}
\hline
\hline
Association	&	Target	&	$R_{\rm cav}$	&	$M_{d}$	&	$d$	&	SpT	&	$M_*$	&	$\log \dot{M}$	&	$L_*$	&	Age$^b$			&	SED?$^c$ & Ref	&		\\
	&		&	(au)	&	($M_{\oplus}$)	&	(pc)	&		&	($M_{\odot}$)	&	($M_{\odot}$	&	($L_{\odot}$)	&	(Myr)			&	\\
 &&&&&&& yr$^{-1}$)&&&\\
\hline
Cham	&	CSCha	&	37	&	29	&	176	&	K2	&	1.5	&	-8.3	&	1.9	&	2	$\pm$	0.5	&	Y	&	1	&		\\
Cham	&	HD97048	&	63	&	322	&	185	&	A0	&	2.4	&	-8.2	&	30	&	4.4	$\pm$	1.1	&	Y	&	1	&		\\
Cham	&	SYCha	&	35	&	16	&	183	&	K7	&	0.8	&	-9.4	&	0.5	&	2	$\pm$	0.5	&	NS	&	2	&		\\
Cham	&	SZCha	&	72$^a$	&	47	&	190	&	K2	&	1.5	&	-7.8	&	1.2	&	2	$\pm$	0.5	&	Y	&	3	&		\\
Cham	&	HPCha	&	50	&	32	&	160	&	K0	&	1.4	&	-9	&	2.4	&	2	$\pm$	0.5	&	NS	&	1	&		\\
CrA	&	PDS99	&	56	&	23	&	155	&	K6	&	0.9	&	-	&	1.1	&	5.5	$\pm$	0.5	&	NS	&	1	&		\\
CrA	&	RXJ1842.9-3532	&	37	&	12	&	154	&	K2	&	0.9	&	-8.8	&	0.8	&	5.5	$\pm$	0.5	&	Y	&	1	&		\\
CrA	&	RXJ1852.3-3700	&	49	&	13	&	146	&	K2	&	1	&	-8.7	&	0.6	&	5.5	$\pm$	0.5	&	Y	&	1	&		\\
EpsiCha	&	TCha 	&	34	&	13	&	110	&	K0	&	1.2	&	-8.4	&	1.3	&	5	$\pm$	3	&	Y	&	1	&		\\
Isolated	&	HD169142	&	24	&	27	&	114	&	A5	&	2	&	-8.7	&	8	&	9	$\pm$	5	&	Y	&	1	&		\\
Isolated	&	V4046Sgr	&	31	&	14	&	72	&	K7	&	0.8	&	-9.3	&	0.5	&	12	$\pm$	2	&	NS	&	1	&		\\
LowerCen	&	HD100453	&	30	&	20	&	104	&	A9	&	1.5	&	<-8.3	&	6.2	&	6.5	$\pm$	0.5	&	Y	&	1	&		\\
LowerCen	&	HD100546	&	25	&	48	&	110	&	B9	&	2.2	&	-7	&	25	&	5.5	$\pm$	1.4	&	Y	&	1	&		\\
Lupus	&	J16070384-3911113	&	52	&	1	&	159	&	M4.5	&	0.2	&	-12.5	&	0	&	2.5	$\pm$	0.5	&	N	&	4	&		\\
Lupus	&	J16070854-3914075	&	40	&	10	&	160	&	-	&	-	&	-	&	-	&	2.5	$\pm$	0.5	&	N	&	4	&		\\
Lupus	&	J16083070-3828268	&	62	&	10	&	156	&	K2	&	1.5	&	-9.1	&	1.8	&	2.5	$\pm$	0.5	&	Y	&	4	&		\\
Lupus	&	J16090141-3925119	&	64	&	1	&	164	&	M4	&	0.2	&	-9.6	&	0.2	&	2.5	$\pm$	0.5	&	N	&	4	&		\\
Lupus	&	RYLup	&	67	&	23	&	159	&	K2	&	1.5	&	-8.2	&	1.9	&	2.5	$\pm$	0.5	&	N	&	4	&		\\
Lupus	&	Sz100	&	26	&	4	&	137	&	M5.5	&	0.1	&	-9.9	&	0.1	&	2.5	$\pm$	0.5	&	N	&	4	&		\\
Lupus	&	Sz111	&	55	&	16	&	158	&	M1	&	0.5	&	-9.6	&	0.2	&	2.5	$\pm$	0.5	&	Y	&	4	&		\\
Lupus	&	Sz118	&	64	&	7	&	164	&	K5	&	1	&	-9.2	&	0.7	&	2.5	$\pm$	0.5	&	N	&	4	&		\\
Lupus	&	Sz123A	&	39	&	4	&	163	&	M1	&	0.6	&	-9.2	&	0.1	&	2.5	$\pm$	0.5	&	N	&	4	&		\\
Lupus	&	Sz91	&	67	&	3	&	159	&	M1	&	0.5	&	-8.7	&	0.2	&	2.5	$\pm$	0.5	&	Y	&	4	&		\\
Oph	&	DoAr44	&	40	&	19	&	146	&	K3	&	1	&	-8.2	&	1.9	&	1.5	$\pm$	0.5	&	Y	&	1	&		\\
Oph	&	IRAS16201-2410	&	32$^a$	&	12	&	157	&	M	&	-	&	-	&	-	&	1.5	$\pm$	0.5	&	Y	&	2	&		\\
Oph	&	IRS48	&	70	&	13	&	134	&	A0	&	2	&	-8.4	&	17.8	&	1.5	$\pm$	0.5	&	Y	&	5	&		\\
Oph	&	ROXRA3	&	49	&	16	&	144	&	M0	&	0.5	&	-8.5	&	0.7	&	1.5	$\pm$	0.5	&	N	&	6	&		\\
Oph	&	RXJ1615.3-3255	&	17$^a$	&	65	&	155	&	K7	&	1.2	&	-8.5	&	0.9	&	1.5	$\pm$	0.5	&	Y	&	7	&		\\
Oph	&	RXJ1633.9-2442	&	36	&	17	&	141	&	K7	&	0.8	&	-10	&	1	&	1.5	$\pm$	0.5	&	Y	&	6	&		\\
Oph	&	SR21	&	56	&	19	&	138	&	G3	&	2	&	-7.9	&	11	&	1.5	$\pm$	0.5	&	Y	&	1	&		\\
Oph	&	SR24S	&	35	&	24	&	114	&	K1	&	1.5	&	-7.2	&	2.5	&	1.5	$\pm$	0.5	&	N	&	1	&		\\
Oph	&	WSB60	&	32	&	19	&	137	&	M6	&	0.2	&	-8.9	&	0.2	&	1.5	$\pm$	0.5	&	Y	&	1	&		\\
Oph	&	WSB82	&	50	&	49	&	156	&	-	&	-	&	-	&	5.1	&	1.5	$\pm$	0.5	&	N	&	6	&		\\
Taurus	&	AATau	&	44	&	12	&	129	&	M0.6	&	0.5	&	-8.4	&	1.1	&	1.5	$\pm$	0.5	&	N	&	1	&		\\
Taurus	&	ABAur	&	156	&	29	&	163	&	A0	&	2.2	&	-6.8	&	65	&	4	$\pm$	1.4	&	Y	&	1	&		\\
Taurus	&	CIDA1	&	21	&	3	&	135	&	M4.5	&	0.2	&	-7.9	&	0.2	&	1.5	$\pm$	0.5	&	N	&	8	&		\\
Taurus	&	CIDA9A	&	29	&	11	&	171	&	M1.8	&	0.4	&	-	&	0.1	&	1.5	$\pm$	0.5	&	NS	&	1	&		\\
Taurus	&	CQTau	&	50	&	41	&	163	&	F2	&	1.6	&	-7	&	10	&	8.9	$\pm$	2.8	&	Y	&	1	&		\\
Taurus	&	DMTau	&	18	&	20	&	145	&	M3.0	&	0.3	&	-8.3	&	0.2	&	1.5	$\pm$	0.5	&	Y	&	1	&		\\
Taurus	&	GGTau	&	224	&	191	&	149	&	K7	&	0.7	&	-7.3	&	1.6	&	1.5	$\pm$	0.5	&	NS	&	1	&		\\
Taurus	&	GMAur	&	40	&	43	&	160	&	K5	&	1	&	-8.3	&	1	&	1.5	$\pm$	0.5	&	Y	&	1	&		\\
Taurus	&	IPTau	&	25	&	3	&	130	&	M0.6	&	0.5	&	-8.1	&	0.6	&	1.5	$\pm$	0.5	&	Y	&	1	&		\\
Taurus	&	LkCa15	&	76	&	33	&	158	&	K5.5	&	0.8	&	-8.4	&	1.3	&	1.5	$\pm$	0.5	&	Y	&	1	&		\\
Taurus	&	MHO2	&	28	&	40	&	133	&	M3	&	0.4	&	-	&	1	&	1.5	$\pm$	0.5	&	N	&	1	&		\\
Taurus	&	MWC758	&	62	&	23	&	160	&	A7	&	1.6	&	-7.4	&	14	&	8.3	$\pm$	0.5	&	Y	&	1	&		\\
Taurus	&	RYTau	&	27	&	39	&	133	&	G0	&	1.7	&	-7.1	&	15	&	1.5	$\pm$	0.5	&	N	&	1	&		\\
Taurus	&	UXTauA	&	31	&	16	&	139	&	G8	&	2.3	&	-8	&	2.5	&	1.5	$\pm$	0.5	&	Y	&	1	&		\\
Taurus	&	V892Tau	&	29$^a$	&	41	&	118	&	B8	&	2	&	-	&	2	&	1.5	$\pm$	0.5	&	Y	&	3	&		\\
Taurus	&	ZZTauIRS	&	58	&	19	&	131	&	M4.5	&	0.2	&	-8.9	&	0.2	&	1.5	$\pm$	0.5	&	N	&	9	&		\\
UpperCen	&	HD135344B	&	52	&	49	&	136	&	F5	&	1.6	&	-7.4	&	6.7	&	5.7	$\pm$	2.8	&	Y	&	1	&		\\
UpperCen	&	HD142527	&	185	&	306	&	157	&	F6	&	2.3	&	-7.5	&	9.9	&	6.6	$\pm$	0.3	&	Y	&	1	&		\\
UpperCen	&	PDS70	&	74	&	12	&	113	&	K7	&	0.8	&	-10.2	&	0.3	&	5	$\pm$	1	&	NS	&	1	&		\\
UpperSco	&	HD143006	&	41	&	20	&	166	&	K0	&	1.6	&	-8.1	&	3.8	&	3.8	$\pm$	1.7	&	Y	&	10	&		\\
UpperSco	&	J160421.7-213028	&	87	&	21	&	150	&	K3	&	1.2	&	-10.5	&	0.7	&	10	$\pm$	0.5	&	Y	&	1	&		\\
\hline
\hline
\end{tabular}}
References. 1) \citet{Francis2020}; 2) \citet{Orihara2022}; 3) \citet{Pinilla2018tds}; 4) \citet{vanderMarel2022}; 5) \citet{vanderMarel2015vla}; 6) \citet{Cieza2020}; 7) \citet{vanderMarel2015-12co}; 8) \citet{Kurtovic2021}; 9) \citet{Hashimoto2021b}; 10) \citet{Huang2018}.\\
$^a$ Cavity size corrected for new Gaia DR2 distance. \\
$^b$ Average age of the cluster following \citet{Michel2021}, individual ages for the Herbig stars from \citet{Vioque2018}.
$^c$ Indicator whether the SED was already indicating the presence of an inner cavity: Y = yes, N = no, NS = no/insufficient \emph{Spitzer} photometry available to make the ass. \\
\end{table}

\begin{table}[!ht]
\caption{Resolved transition disks within 200 pc with cavities $<$15 au}
\label{tbl:smallsample}
\resizebox{\textwidth}{!}{%
\begin{tabular}{lllllllll}
\hline
\hline
Association	&	Target	&	$R_{\rm cav}$	&	$d$	&	SpT	&	$M_*$	&	$\dot{M}$	&	SED?$^a$ & Ref \\
&&(au)&(pc)&&($M_{\odot}$)&($M_{\odot}$ yr$^{-1}$)& \\
\hline
Lupus	&	Sz129	&	10	&	162	&	K7	&	0.78	&	-8.4	&	N	&	1\\
Lupus	&	Sz76	&	4	&	160	&	M4	&	0.23	&	-9.3	&	Y	&	1\\
Lupus	&	J16102955-3922144	&	11	&	160	&	M4.5	&	0.22	&	-9.8	&	Y	&	1\\
Lupus	&	Sz84	&	12	&	153	&	M5	&	0.17	&	-9.23	&	Y	&	1\\
Taurus	&	UZTauE	&	11	&	134	&	M1	&	0.89	&	-	&	N	&	2\\
Taurus	&	MHO6	&	10	&	141	&	M5	&	0.12	&	-9.3	&	NS	&	3\\
Taurus & XZTauB$^b$ & 1.3 & 146 & M2 & 0.37 & -7.9 & N & 4 \\
TWHya	&	TWHya	&	3	&	60	&	M0	&	0.6	&	-8.9	&	Y	&	5\\
Oph & ISO-Oph2B & 2.2 & 134 & - & - & - & N & 6 \\ 
\hline
\hline
\end{tabular}}
References. 1) \citet{vanderMarel2022}; 2) \citet{Long2019}; 3) \citet{Kurtovic2021}; 4) \citet{Osorio2016}; 5) \citet{Andrews2016}; 6) \citet{Gonzalez2020}; \\
$^a$ Indicator whether the SED was already indicating the presence of an inner cavity: Y = yes, N = no, NS = no/insufficient \emph{Spitzer} photometry available to make the ass. \\
$^b$ Tentative: not recovered in \citet{Ichikawa2021}. \\
\end{table}

\begin{table}[!ht]
\caption{Limits on (sub)stellar companions in transition disks}
\label{tbl:circumbinary}
\resizebox{\textwidth}{!}{%
\begin{tabular}{lllll}
\hline
\multicolumn{5}{c}{Circumbinary disks in transition disk sample} \\
\hline
Target & $R_{\rm cav}$ & Binary sep. & Binary & Ref \\
& (au) & (au) &sufficient? & \\
\hline
CS~Cha &37&$\sim4$ (RV)&N?&1 \\ 
RXJ1633.9-2442$^*$ & 36 & 3.3 & N? & 12 (not detected in 7) \\
DM~Tau$^*$ & 18 & 6.5 & Y? & 2\\
GG~Tau &224&35&Y& 3 \\
HD142527 &185&13&Y&4 \\
RY~Tau&27&? (RV)&?&1 \\
V892~Tau &34&8&Y&5 \\
V4046Sgr &31&0.045 & N & 6\\
\hline
\multicolumn{5}{c}{Transition disks for which a inner (sub)stellar companion ($>20 M_{\rm Jup}$) has been excluded} \\
\hline
Target&$R_{\rm cav}$ & Range & & Ref \\
&(au)&(au)&& \\
\hline
DoAr44 & 40 & 1-47 & & 7\\
SR21 & 56 & 1-44 && 7,12 \\
SR24S & 35 & 1-36 && 7,12 \\
RXJ1604.3-2130 & 87 & 1-48 && 8 \\
CIDA9 & 29 & 1-55 && 9\\
DM~Tau & 18 & 1-46 && 9,10 \\
GM~Aur & 40 & 1-51 && 9,10 \\
IP~Tau & 25 & 1-42 && 9\\
LkCa15 & 76 & 1-15 && 9,10 \\
RY~Tau & 27 & 1-43 && 9,10 \\
UX~Tau A & 31 & 1-45 && 9,10\\
ZZ~Tau IRS & 58 & 1-42 && 9 \\
RXJ1615.3-3255 & 17 & 2-31 && 2 \\
RXJ1842.9-3532 & 37 & 2-31 && 2 \\
\hline
\multicolumn{5}{c}{Transition disks for which a spectroscopic binary has been excluded} \\
\hline
Target&$R_{\rm cav}$ & & & Ref \\
&(au)&& \\
\hline
SZ~Cha & 72 &&& 11 \\
RXJ1852.3-3700 & 49&&& 11 \\
SR21 &56 &&& 11 \\
IRAS~16201-2410 &32&&& 11 \\
RXJ1633.9-2442 &36&&& 11 \\

\hline
\hline
\end{tabular}} \\
$^*$ Tentative\\
References 1. \citet{Nguyen2012}; 2) \citet{Willson2016}; 3) \citet{White1999}; 4) \citet{Lacour2016}; 5) \citet{Leinert1997}; 6) \citet{Stempels2004}; 7) \citet{Cheetham2015}; 8) \citet{Kraus2008}; 9) \citet{Kraus2011}; 10) \citet{Pott2010}; 11) \citet{Kohn2016}; 12) \citet{Ruiz2016}.
\end{table}





\end{appendices}


\section{Data availability statement}
All data generated or analysed during this study are included in this published article (and its supplementary information files).

\clearpage
\newpage

\bibliography{myrefs}
\bibliographystyle{sn-basic}



\end{document}